\documentclass[useAMS,usenatbib]{mn2e}

\usepackage[psamsfonts]{amssymb}
\usepackage[dvips]{graphicx}
\usepackage{amsmath,alltt}
\usepackage{multirow}
\usepackage{rotating}
\usepackage{lscape}

\title[The 4-Body Problem in Newtonian gravity I]{The chaotic four-body problem in Newtonian gravity I: Identical point-particles}
\author[Leigh N. W. C., Stone, N. C., Geller A. M., Shara M. M., Muddu H., Solano-Oropeza D., Thomas Y.]{Nathan W. C. Leigh$^{1}$, Nicholas C. Stone$^{2,3}$, Aaron M. Geller$^{4,5}$, Michael M. Shara$^{1}$, 
\newauthor
Harsha Muddu$^{6}$, Diana Solano-Oropeza$^{6}$, Yancey Thomas$^{6}$
\thanks{E-mail: nleigh@amnh.org (NWCL)}\\
$^{1}$Department of Astrophysics, American Museum of Natural History, Central Park West and 79th Street, New York, NY 10024 \\
$^{2}$Columbia Astrophysics Laboratory, Columbia University, New York, NY, 10027, USA \\
$^{3}$Einstein Fellow \\
$^{4}$Center for Interdisciplinary Exploration and Research in Astrophysics (CIERA) and Department of Physics and Astronomy, \\ Northwestern University, 2145 Sheridan Rd, Evanston, IL 60208, USA \\
$^{5}$Adler Planetarium, Dept.\ of Astronomy, 1300 S. Lake Shore Drive, Chicago, IL 60605, USA \\
$^{6}$Student Research and Mentoring Program, Department of Education, American Museum of Natural History, \\
Central Park West and 79th Street, New York, NY 10024}
\begin{document}

\pagerange{\pageref{firstpage}--\pageref{lastpage}} \pubyear{2016}

\maketitle

\label{firstpage}

\begin{abstract}
In this paper, we study the chaotic four-body problem in Newtonian gravity.  Assuming point particles and total 
encounter energies $\le$ 0, the 
problem has three possible outcomes.  
We describe each outcome as a series of discrete transformations in energy space, 
using the diagrams first presented in Leigh \& Geller (2012; see the Appendix).  
Furthermore, we develop a formalism for calculating probabilities for these outcomes 
to occur, expressed using the density of escape configurations per unit energy, and based on the Monaghan 
description originally developed for the three-body problem.  
We compare this analytic formalism to results from a series of binary-binary encounters with identical point particles,
simulated using the \texttt{FEWBODY} code.  Each of our three encounter outcomes produces a unique 
velocity distribution for the escaping star(s).  Thus, these distributions can potentially 
be used to constrain the origins of dynamically-formed populations, via a 
direct comparison between the predicted and observed velocity distributions.  Finally, we show that, for encounters 
that form stable triples, the simulated single star escape velocity distributions are the same as for the 
three-body problem.  This is also the case for the other two encounter outcomes, but only at low virial ratios.  
This suggests that single and binary stars processed via single-binary and binary-binary encounters in 
dense star clusters should have a unique velocity distribution relative to the underlying Maxwellian distribution 
(provided the relaxation time is sufficiently long), which can be calculated analytically.
 
\end{abstract}


\section{Introduction} \label{intro}


The $N$-body problem is a longstanding issue hailing from Sir Isaac Newton's day \citep{newton1686}, as discussed in a previous paper 
in this series \citep{leigh15,valtonen06}.  It has been the subject of intense study for centuries, with a flurry of rapid progress over the last few decades due to the introduction of  computers.  
And yet, small-$N$ chaos is an essential puzzle related to dynamical phenomena such as direct stellar and binary interactions in globular \citep[e.g.,][]{heggie75,leigh12,leigh13b}, open \citep[e.g.,][]{leigh11,leigh13a,geller15,leigh16} and even nuclear \citep[e.g.][]{davies98,merritt13} star clusters.  The three-body, and more generally the $N$-body, problem has never been fully solved analytically; instead, approximate methods are used along with simplifying assumptions that make the problem tractable.  Examples include simple three-body configurations such as Burrau's Pythagorean problem \citep{burrau1913}, the restricted three-body problem often applied to the Earth-Sun-Moon system, periodic solutions to the non-hierarchical three-body problem \citep[e.g.][]{suvakov13} and finally complex
numerical simulations of very large-$N$ systems. These $N$-body simulations can be prohibitively computationally expensive, with integration times scaling with the number of particles as $N^2$.  Furthermore, star cluster simulations that include both stable and chaotically-interacting multiple star systems can significantly slow the integration times down, since the need for small or short time-steps during these small-$N$ interactions occurring within the larger framework of the cluster simulation can contribute to a significant increase in the overall computer run-times of the simulations \citep{hurley05,valtonen06,geller13}.  This is particularly problematic, since simulations 
have shown that encounters involving binaries and triples can be crucial for not only the overall cluster 
evolution \citep[e.g.][]{hut83a,hut92}, but also for the formation of exotic populations such as blue stragglers 
\citep{perets09,leigh11,geller13,naoz14} and even accreting \citep{mapelli14} or massive \citep{stone16} black holes.  What's more, observations 
have now revealed that higher-order multiple star systems are present in young star clusters in 
non-negligible numbers \citep[e.g.][]{leigh13a}.

We argue here that $N =$ 4 is an optimal number for studying the chaos of gravitationally-interacting particles, with an emphasis on going beyond the already well-studied three-body problem.  This is because $N =$ 4 offers a reasonable balance between the computer run-times for the simulations, and the statistical significance of our analysis, which depends directly on the total number of simulations performed for a given set of initial conditions.  Theoretically, binary-binary encounters should dominate over single-binary encounters in any star cluster with a binary fraction $f_{\rm b} \gtrsim$ 10\% \citep{sigurdsson93,leigh11}.  For these reasons, we focus on the chaotic $N =$ 4 problem in this paper.  We further narrow our focus to consider only identical, equal-mass point particles.  Regardless, in Section~\ref{discussion}, we use our results to predict the expected behavior of interactions involving non-identical particles with different masses, which we intend to test directly in future work.

The three-body system with $N = 3$ typically evolves via a series of close triple encounters  \citep[e.g.][]{agekyan67,anosova94}.  Between each such event, one of the objects is temporarily ejected but remains bound to the three-body system.\footnote{We will use the term "ejection" to refer to these types of events, where the ejected object actually remains bound to the system.  We will use the term "escape" to refer to events where particles become unbound from the remaining system.}  This object recoils some distance from the remaining binary before returning to initiate another triple encounter.  Eventually, one of the bodies is ejected with a sufficiently high velocity to become unbound, and it escapes to infinity.  This chaotic progression or evolution can be simplified as follows.  

The time evolution of the chaotic $N = 3$ problem can be broken down into a single discrete transformation, which occurs in energy and angular momentum space.  The initial conditions for the encounter define the relative energies and angular momenta of the single star and the binary.  After a transformation is applied in energy and angular momentum space, the final state of the system is qualitatively the same as the initial state (i.e., a single star and a binary are left over), but quantitatively the relative energies and angular momenta are distributed differently among the particles in the final state. Hence, the time evolution of the system connecting the initial and final states can, to first-order, be described as a single transformation in energy and angular momentum space, where the intermediary chaos is neglected such that statistical ensembles of outcomes are more easily considered, rather than individual choices of scattering parameters.  

In this paper, we extend this approach to the four-body problem which, as we will show, can similarly be broken down into a series of discrete transformations in energy space (and angular momentum space).  In the case of the four-body problem with point particles, the transformations lead to one of three final states for the system (instead of one final state in the 
three-body problem).
We use the numerical scattering code \texttt{FEWBODY} to simulate a series of binary-binary encounters involving identical point particles, for different values of the virial ratio.
In Section~\ref{method}, we describe the simulations used in this study, and present the 
resulting distributions of final virial ratios, escape velocities and encounter times.  We go on to apply the time-averaged 
virial approximation to 
better understand the response of the interacting system to particle escapes in Section~\ref{model}, and adapt the Monaghan 
formalism, originally derived by \citet{monaghan76a} for the three-body problem, to describe the distributions of single 
star escape velocities.  The significance of our results for astrophysical observations are discussed in Section~\ref{discussion}, 
and our key results are summarized in Section~\ref{summary}.
 
\section{Method} \label{method}

In this section, we present the numerical scattering experiments used to study the time evolution of 
the chaotic 4-body problem as a function of the initial virial ratio.  

\subsection{Numerical scattering experiments} \label{exp}

We calculate the outcomes of a series of binary-binary (2+2) encounters using the \texttt{FEWBODY} numerical 
scattering code\footnote{For the source code, see http://fewbody.sourceforge.net.}.  The code integrates the usual 
$N$-body equations in configuration- (i.e., position-) space in order to advance the system forward in time, using the 
eighth-order Runge-Kutta Prince-Dormand integration method with ninth-order error estimate and adaptive time-step.  
For more details about the \texttt{FEWBODY} code, we refer the reader to \citet{fregeau04}.  

The outcomes of these 2+2 encounters are studied as a function of the initial virial ratio $k$, defined as:
\begin{equation}
\label{eqn:virial}
k = \frac{T_{\rm 1} + T_{\rm 2}}{E_{\rm b,1} + E_{\rm b,2}},
\end{equation}
where the indexes 1 and 2 correspond to the two initial binaries.  The initial kinetic energy corresponding to the 
centre of mass motion of binary $i$ is:
\begin{equation}
\label{eqn:kinetic}
T_{\rm i} = \frac{1}{2}m_{\rm i}v_{\rm inf,i}^2,
 \end{equation}
where $m_{\rm i} =$ $m_{\rm i,a} + m_{\rm i,b}$ is the total binary mass and $v_{\rm inf,i}$ is the initial 
centre of mass velocity for binary $i$.  The initial orbital energy of binary $i$ is:
\begin{equation}
\label{eqn:orbital} 
E_{\rm b,i} = -\frac{Gm_{\rm i,a}m_{\rm i,b}}{2a_{\rm i}},
\end{equation}
where $m_{\rm i,a}$ and $m_{\rm i,b}$ are the masses of the binary components and $a_{\rm i}$ is the initial 
orbital separation.  Given this definition for the virial ratio, $k =$ 0 corresponds to the binaries starting from rest and 
$k =$ 1 corresponds to a relative velocity equal to the critical velocity $v_{\rm crit}$, defined as the relative velocity at infinity 
needed for a total encounter energy of zero.  That is, Equation~\ref{eqn:virial} can be re-written as:
\begin{equation}
\label{eqn:virial2}
k = \Big( \frac{v_{\rm rel}}{v_{\rm crit}} \Big)^2,
\end{equation}
where $v_{\rm rel}$ is the initial relative velocity at infinity between the binaries and $v_{\rm crit}$ the critical velocity.  We consider 
initial virial ratios of $k =$ 0.00, 0.04, 0.16, 0.36, 0.64 and 1.00.

All objects are point particles with masses of 1 M$_{\odot}$.  All binaries have $a_{\rm i} =$ 1 AU initially, and 
eccentricities $e_{\rm i} =$ 0.  We fix the impact parameter at $b =$ 0 for all simulations.  
The angles defining the initial relative configurations of the binary orbital planes and phases are 
chosen at random.  We perform 10$^4$ numerical scattering experiments for every initial virial ratio.  For comparison purposes, 
we also run one set of 10$^4$ simulations at $k = 0.04$ with identical initial conditions, but assuming initial binary orbital 
separations of 1 AU and 10 AU.  These simulations are shown by the solid triangles in Figure~\ref{fig:fig4} and are 
compared to our fiducial set of simulations for two initially 1 AU binaries in Figure~\ref{fig:fig9}.

We use the same criteria as \citet{fregeau04} to decide when a given encounter is complete.  To first order, this is defined as 
the point at which the separately bound hierarchies that make up the system are no longer interacting with each other or 
evolving internally.  More specifically, the integration is terminated when the top-level hierarchies have positive relative 
velocity and the corresponding top-level $N$-body system has positive total energy.  Each hierarchy must also be dynamically 
stable and experience a tidal perturbation from other nodes within the same hierarchy that is less than the critical value 
adopted by \texttt{FEWBODY}, called the tidal tolerance parameter.  For this study, we adopt the a tidal tolerance parameter 
$\delta =$ 10$^{-10}$ for all simulations.\footnote{The more stringent the tidal tolerance parameter is chosen to be, the closer to a 
"pure" $N$-body code the simulation becomes.}    This choice for $\delta$, while computationally expensive, is needed to maximize 
the accuracy of our simulations, and ensure that we have converged on the correct encounter outcome (see \citealt{geller15} 
for more details).

\subsection{Results}

The chaotic four-body problem involving point particles has three possible outcomes, provided the total encounter 
energy satisfies $E \le$ 0.\footnote{We ignore encounters producing four single stars, since these require positive total 
encounter energies and hence very large relative velocities at infinity.  As explained in Appendix~\ref{appendix}, such encounters are unlikely to occur in dense stellar systems, since dynamically ``softÕÕ binaries dissociate rapidly from the cumulative effects of many weak three-body encounters.  Only the far tail of the Maxwellian will be capable of completely ionizing two hard binaries, and the rate of this will be exponentially suppressed unless both binaries are only marginally hard.}  That is, when an encounter is over, the remaining 
configuration is described by one of the following outcomes:
\begin{list}{*}{} 
\item two binaries (2+2)
\item a triple and a single star (3+1)
\item a binary and two single stars (2+1+1) 
\end{list}
Although complete ionizations, or $1+1+1+1$ outcomes, are technically possible in some scenarios, they are extremely rare 
in any realistic star cluster environment.  We defer a more in depth discussion of this to Appendix~\ref{appendix}.

\begin{figure}
\begin{center}                                                                                                                                                           
\includegraphics[width=\columnwidth]{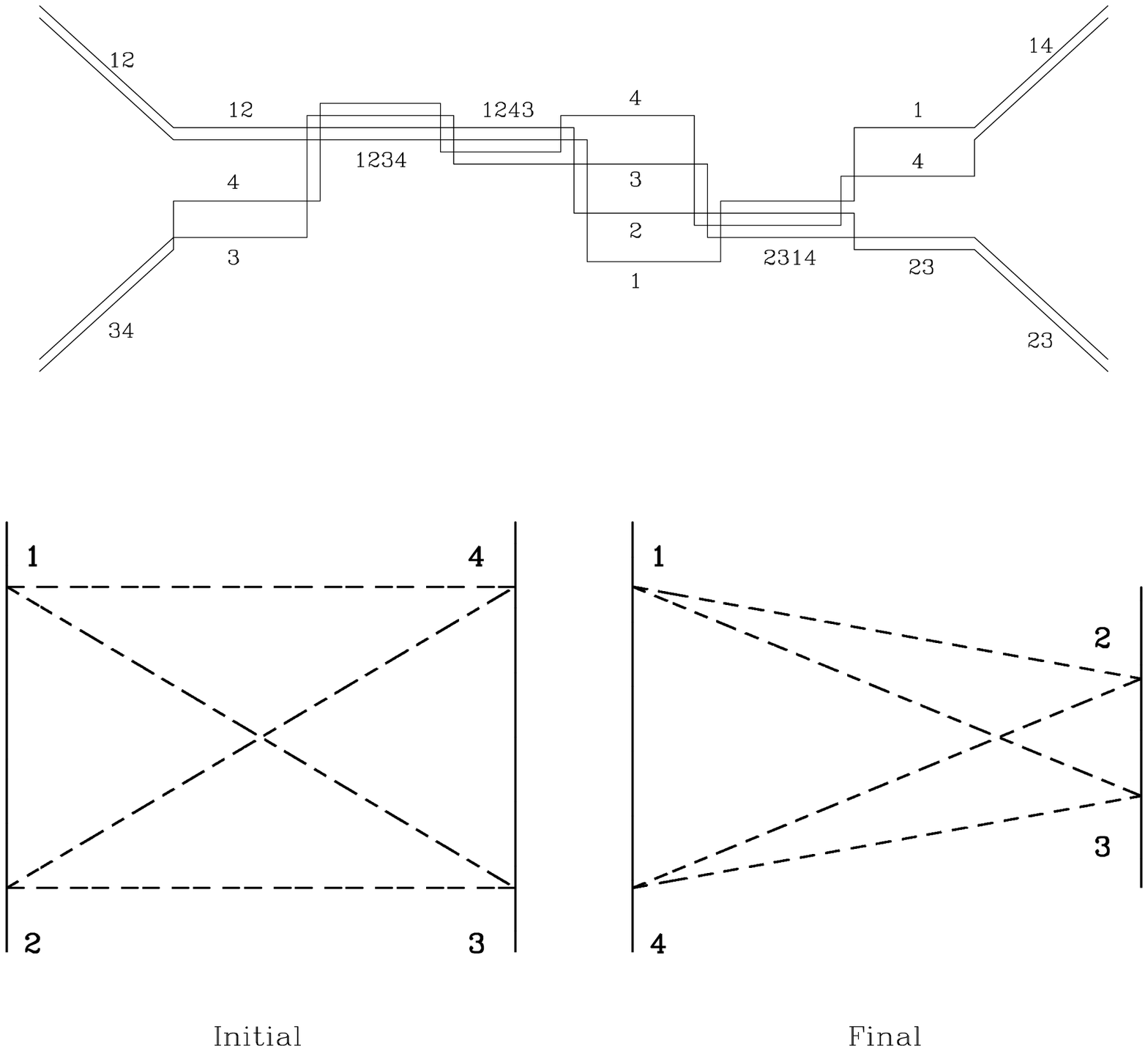}
\end{center}
\caption[Schematic diagrams for the time evolution of a 2+2 encounter producing a 2+2 outcome]{The time evolution of an 
example 2+2 simulation producing a 2+2 outcome is shown schematically.  In the top diagram, time increases from left to 
right along the x-axis, whereas the y-axis shows the progression of configurations.  In the bottom diagram, with two distinct 
insets, shows the initial and final states of the encounter in energy space, using the energy diagrams first introduced in \citet{leigh12}.  Note that most 2+2 outcomes in our simulations produce binaries with a very large difference in their orbital energies, even more exaggerated than shown here.  
\label{fig:fig1}}
\end{figure}

\begin{figure}
\begin{center}                                                                                                                                                           
\includegraphics[width=\columnwidth]{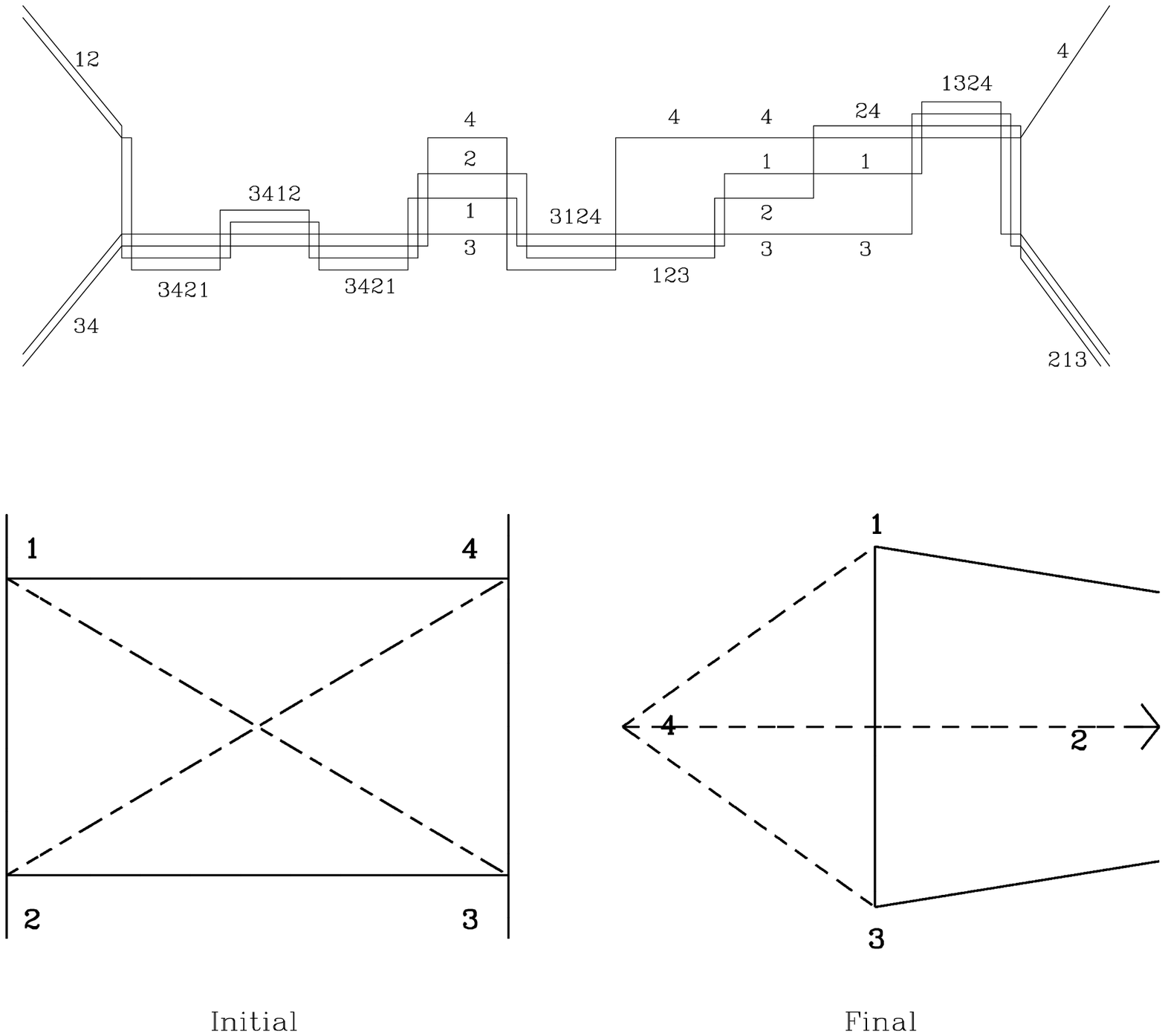}
\end{center}
\caption[Schematic diagrams for the time evolution of a 2+2 encounter producing a 3+1 outcome]{The time evolution of an 
example 2+2 simulation producing a 3+1 outcome is shown schematically.  In the top diagram, time increases from left to 
right along the x-axis, whereas the y-axis shows the progression of configurations.  The bottom diagram, with two distinct 
insets, shows the initial and final states of the encounter in energy space, using the energy diagrams first introduced in 
\citet{leigh12}.  Note that most 3+1 outcomes in our simulations produce a relatively low-velocity escaping single star and a triple with a very large difference in their inner and outer orbital energies, even more exaggerated than shown here.
\label{fig:fig2}}
\end{figure}

\begin{figure}
\begin{center}                                                                                                                                                           
\includegraphics[width=\columnwidth]{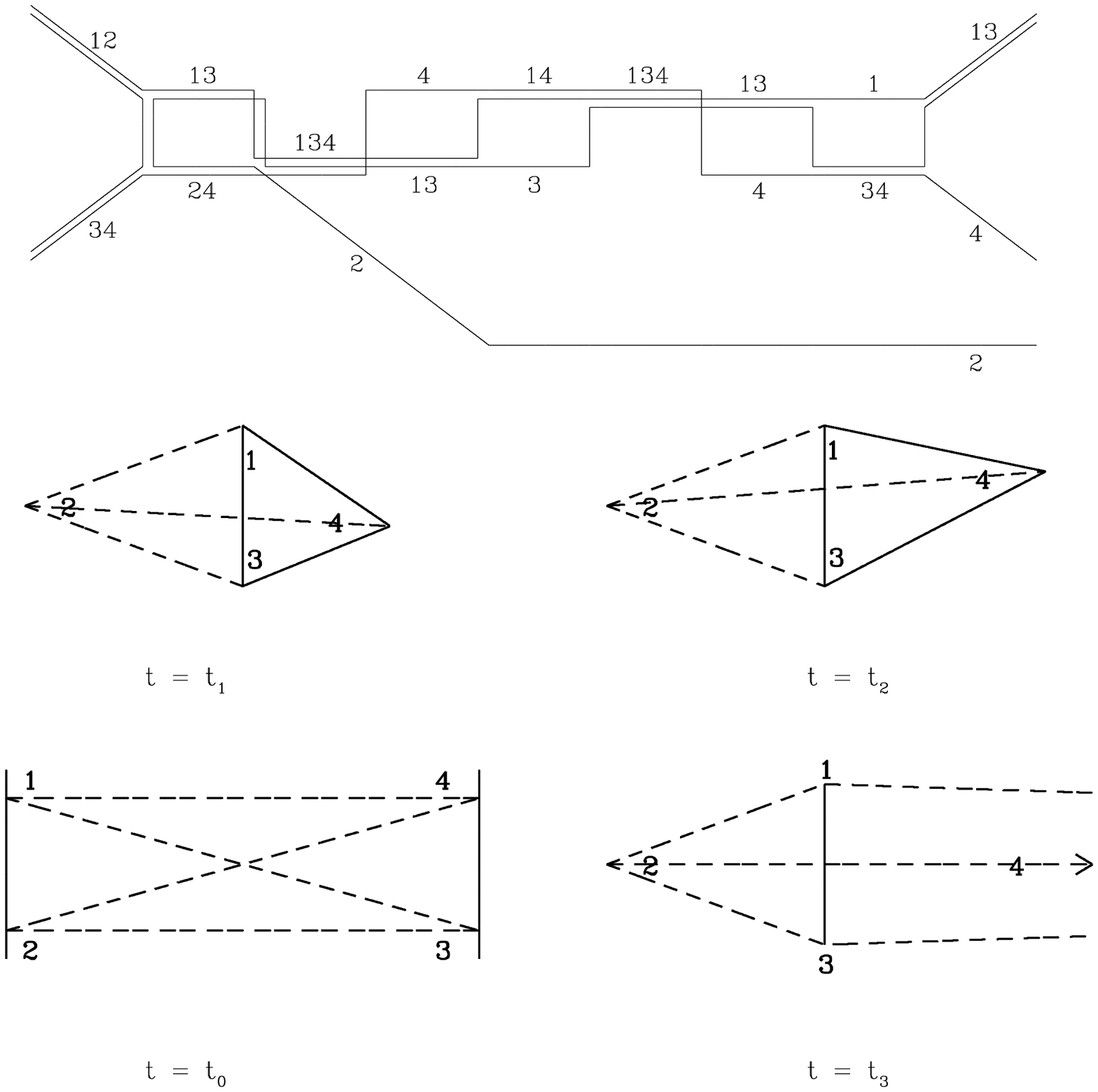}
\end{center}
\caption[Schematic diagrams for the time evolution of a 2+2 encounter producing a 2+1+1 outcome]{The time evolution of an 
example 2+2 simulation producing a 2+1+1 outcome is shown schematically.  In the top diagram, time increases from left to 
right along the x-axis, whereas the y-axis shows the progression of configurations.  The bottom diagram, with four distinct 
insets, we show the time evolution of the encounter in energy space, using the energy diagrams first introduced in \citet{leigh12}.  This 
figure is reproduced from Figures A2 and A3 in \citet{leigh12}.
\label{fig:fig3}}
\end{figure}

The (qualitative) time evolution of three example simulations, each ending in one of the three outcomes listed above, are depicted 
in the top insets of Figures~\ref{fig:fig1}, ~\ref{fig:fig2} and~\ref{fig:fig3} using the schematic diagrams first introduced in \citet{hut83b}.  
The three outcomes are also depicted schematically in the bottom insets of Figures~\ref{fig:fig1}, ~\ref{fig:fig2} and~\ref{fig:fig3} using the energy diagrams first presented in \citet{leigh12}.  The latter diagrams quantify the exchange of energy between particles during an encounter.  Briefly, each angle of the polygon corresponds to the fraction of the total encounter energy contained in star $i$.  The solid 
and dashed lines connect particles that are gravitationally bound and unbound, respectively.  For more details regarding the details of these energy diagrams, see the Appendix in \citet{leigh12}.

The energy diagrams shown in the bottom insets of Figures~\ref{fig:fig1}, ~\ref{fig:fig2} and~\ref{fig:fig3} correspond to nearly instantaneous, or "discrete" events (i.e., close or strong encounters between individual particles); to first order, each time an object passes near 
the centre of mass of the system, its motion is significantly perturbed and it recedes from the centre of mass on a new orbit, before returning to 
repeat the process.  Thus, in energy space, 
the time evolution of the four-body system can be described by a series of discrete transformations in energy space, as depicted schematically in the bottom panels of Figures~\ref{fig:fig1}-~\ref{fig:fig3}.  In principle, this has the potential to simplify the problem, since we are not concerned with the 
evolution in position and velocity space, where long excursions can occur with little to no exchange of energy (or angular momentum) 
between particles.  This same procedure can be applied analogously in angular momentum space.

Importantly, it is the first escape event that decides the outcome of a chaotic 4-body encounter with non-positive total energy; that is, the first object to become gravitationally unbound and leave the system with positive total energy (ignoring the binding energies of any left-over binary orbits) decides the final outcome.  This can be understood as follows.  
If a binary is the first object to escape, then the encounter is over and the outcome is 2+2.  If, on the other hand, a single star is the first object to escape, then the encounter either terminates immediately if the remaining triple is dynamically stable (i.e., the outcome is 3+1), or continues to evolve chaotically until the temporary triple eventually disrupts (i.e., the outcome is 2+1+1).  It is not possible for an initially unstable isolated triple to become stable \citep{littlewood52}.  Thus, the stability of the triple configuration immediately after the escape of the first single star, and, more generally, the first escape event, entirely decides the encounter outcome.  

\begin{figure}
\begin{center}                                                                                                                                                           
\includegraphics[width=\columnwidth]{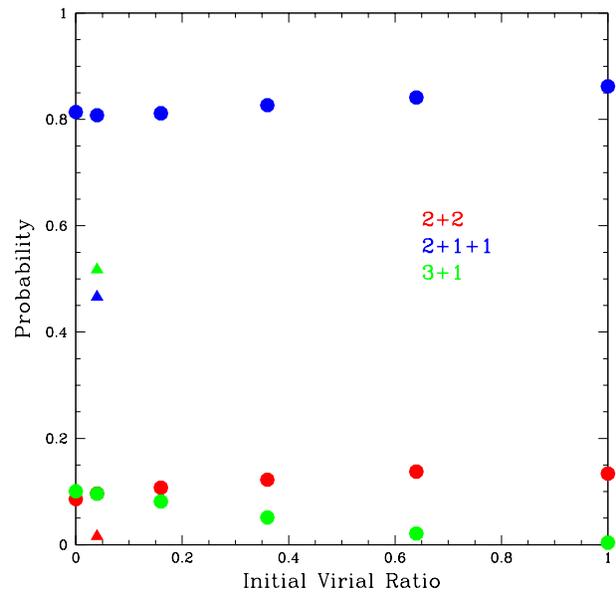}
\end{center}
\caption[Fraction of encounters resulting in each encounter outcome as a function of the initial virial ratio]{The fraction 
of simulations that end in each of our three encounter outcomes are shown as a function of the initial virial ratio.  The blue, red 
and green points correspond to the 2+1+1, 2+2 and 3+1 outcomes, respectively.  For comparison, we also show by the solid triangles these fractions for $k = 0.04$ assuming initial binary separations of $a_{\rm 1} =$1 AU and $a_{\rm 2} =$ 10 AU.
\label{fig:fig4}}
\end{figure}

In Figure~\ref{fig:fig4}, we show the fraction of simulations that end in each of our three encounter outcomes, as a 
function of the initial virial ratio.  Note that these results are sensitive to our choice of particle masses and initial binary energies.  The blue, red and green points correspond to the 2+1+1, 2+2 and 3+1 outcomes, respectively.  Encounters 
producing two single stars dominate for our choice of the initial conditions, giving outcome fractions of $\sim$ 80\%-85\%.  At low virial ratios 
(i.e. $k \sim$ 0), 2+2 and 3+1 are roughly equally probable (i.e., $\sim$ 10\% of the outcomes are 2+2, and $\sim$ 10\% are 3+1).  As 
the virial ratio increases, however, the fraction of 3+1 outcomes slowly decreases reaching $\sim$ 0\% at $k =$ 1, while the fractions of 
2+2 and 2+1+1 outcomes both increase by $\sim$ a few percent.  These results are in good agreement with those presented by \citet{mikkola83}, who performed numerical scattering experiments using very similar initial conditions (i.e., all identical particles, and 
identical binaries).  

\begin{figure*}
\begin{center}                                                                                                                                                           
\includegraphics[width=\textwidth]{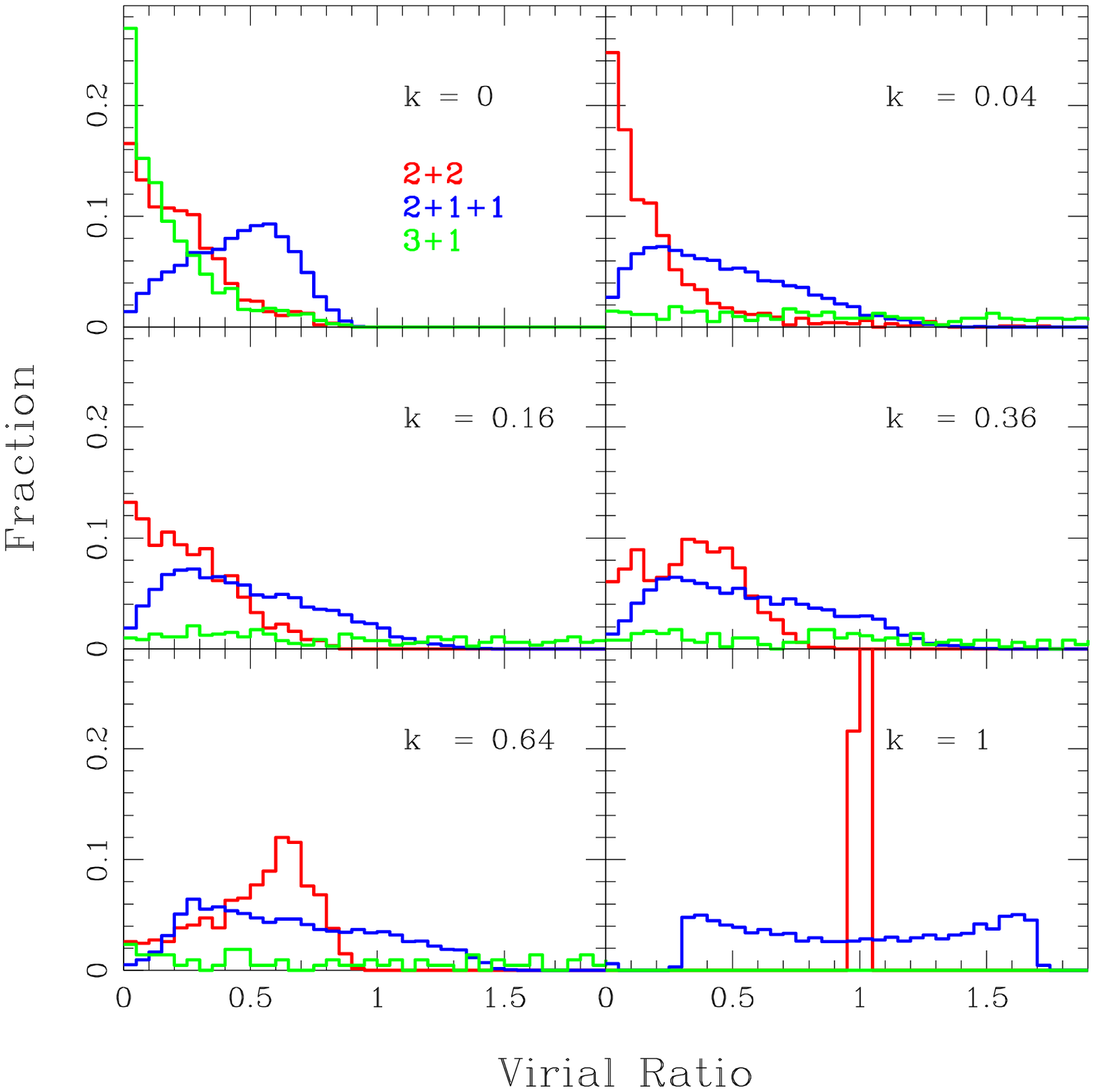}
\end{center}
\caption[Distributions of final virial ratios for every initial virial ratio]{The distributions of final 
virial ratios are shown.  The blue, red 
and green histograms correspond to the 2+1+1, 2+2 and 3+1 outcomes, respectively.  Each panel shows the distributions 
for a different value of the initial virial ratio.  All histograms have been normalized by the total number of simulations that resulted in  the corresponding outcome.
\label{fig:fig5}}
\end{figure*}

\begin{figure*}
\begin{center}                                                                                                                                                           
\includegraphics[width=\textwidth]{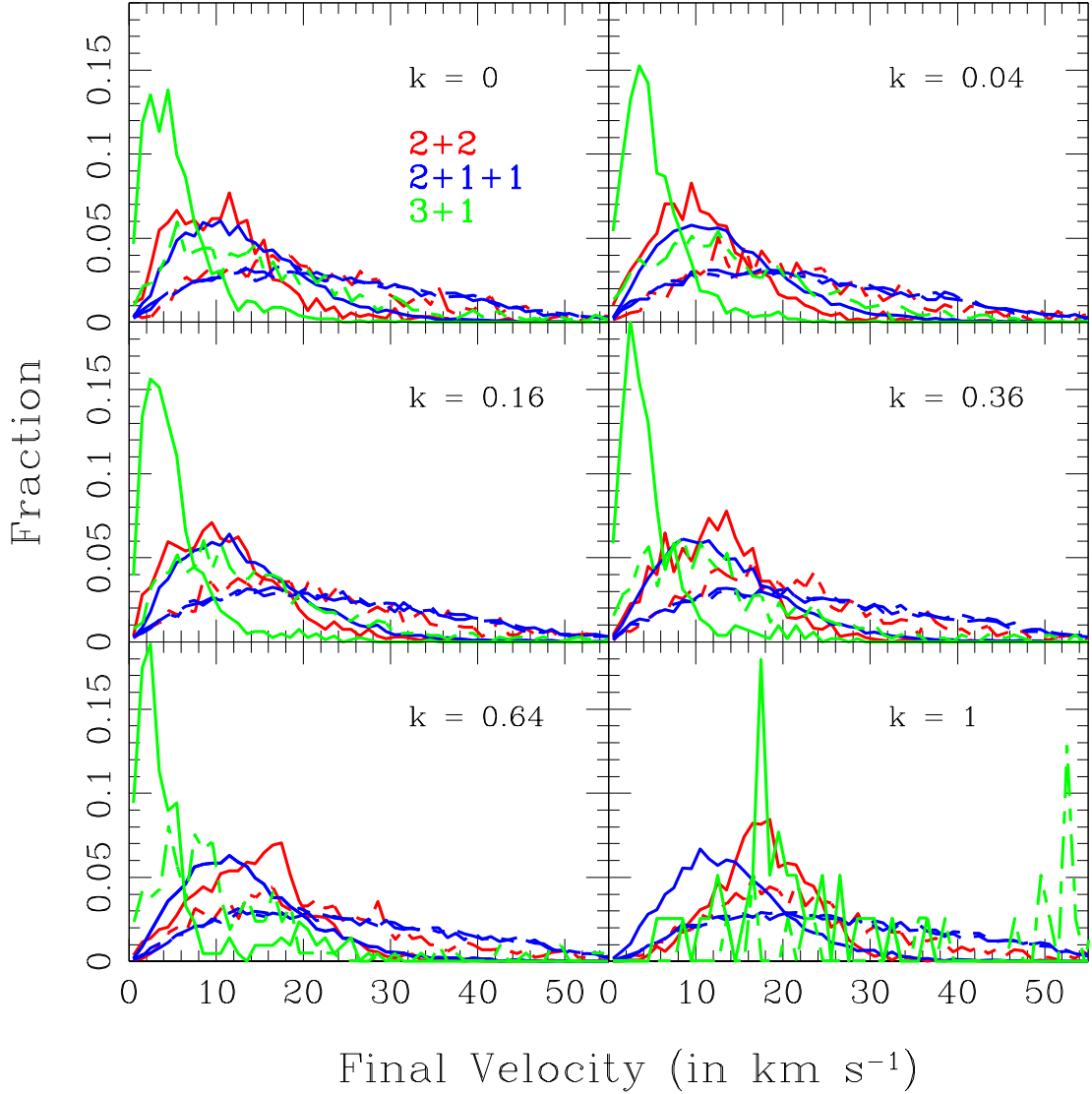}
\end{center}
\caption[Distributions of initial and final escape velocities for every initial virial ratio]{The distributions of initial and final 
escape velocities are shown in km s$^{-1}$.  The blue, red 
and green histograms correspond to the 2+1+1, 2+2 and 3+1 outcomes, respectively.  We use different line types to show the 
distributions for the different objects produced for each outcome; solid lines correspond to multiple star systems (i.e., binaries and triples) and dashed lines correspond to single stars (with the exception of the 2+2 case, for which the dashed red lines simply correspond to the other binary).   Each panel shows the 
distributions for a different value of the initial virial ratio.  All histograms have been normalized by the total number of simulations that resulted in the corresponding outcome.
\label{fig:fig6}}
\end{figure*}

\begin{figure*}
\begin{center}                                                                                                                                                           
\includegraphics[width=\textwidth]{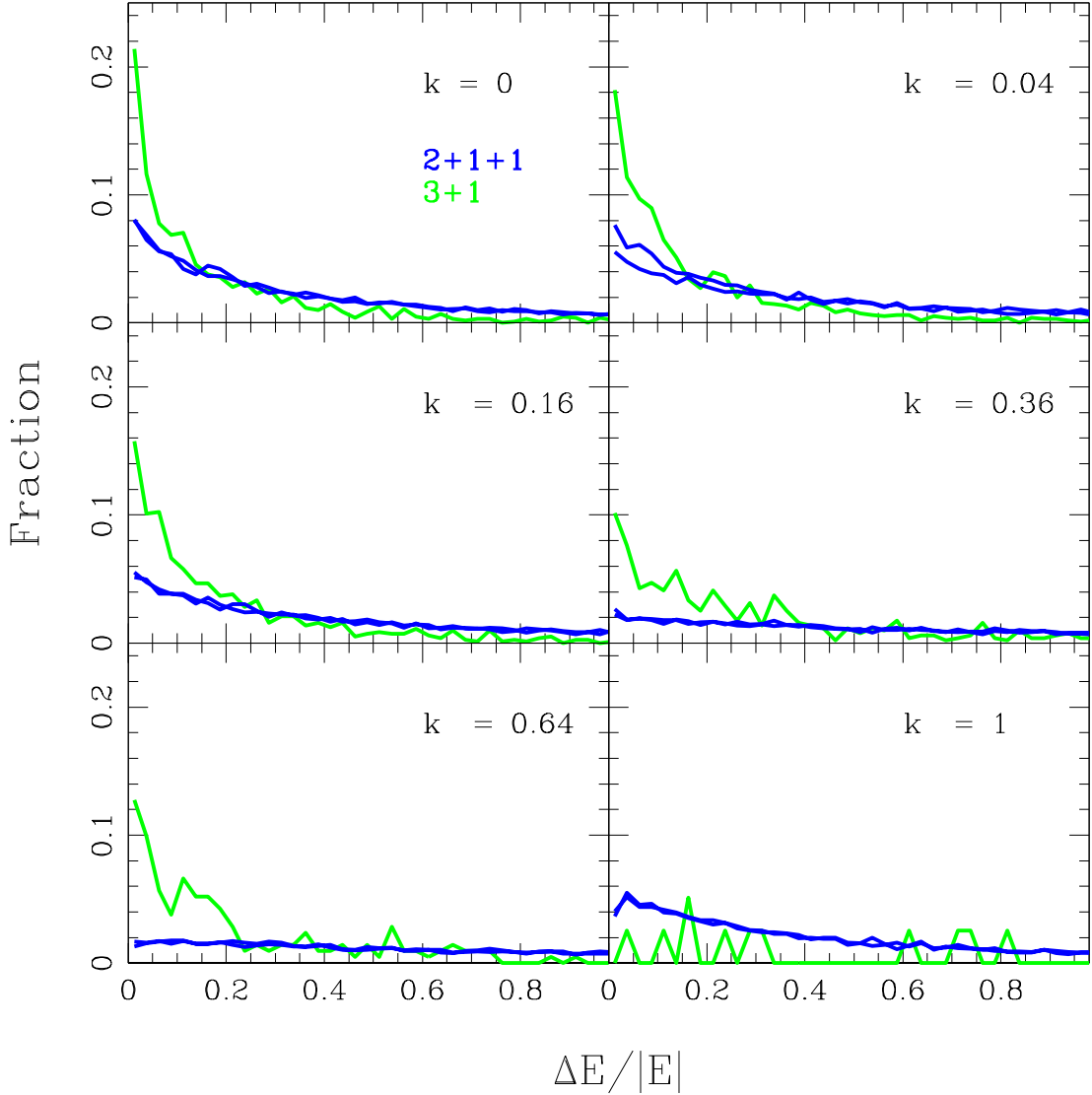}
\end{center}
\caption[Fractional change in energy for all escape events]{The distributions of the fractional change in 
energy are shown for all escape events.  The different panels show the results for different values of the initial virial 
ratio.  The colour coding is the same as in 
Figures~\ref{fig:fig4}-\ref{fig:fig6}.  We plot the fractional change in energy only if $\Delta{E}$/$|E| >$ 0 and, to avoid confusion, omit the 2+2 outcome.  That is, the fractional 
change in energy is plotted for the escaping single star during 3+1 outcomes (green), and for both escaping single stars during 
2+1+1 outcomes (blue).   All histograms have been normalized by the total number of simulations that resulted in the corresponding outcome.
\label{fig:fig7}}
\end{figure*}

In Figure~\ref{fig:fig5} we show the distributions of initial and final virial ratios, and in Figure~\ref{fig:fig6} we show the distributions of initial 
and final velocities (in km s$^{-1}$).  Figure~\ref{fig:fig7} shows the distributions of the fractional change in energy, for all escape events, 
provided they satisfy $\Delta{E}$/$|E| >$ 0.  That is, the fractional 
change in energy is plotted for the escaping single star during 3+1 outcomes (green), and for both escaping single stars during 2+1+1 
outcomes (blue).  Note that the quantity $\Delta{E}$/$|E|$ is positive if the escaping object is a single star, or if the escaping object is a binary and the absolute value of its orbital energy is less than its (translational) kinetic energy.  Hence, to avoid confusion, we omit the 2+2 outcome in Figure~\ref{fig:fig7}, and note that in Figures~\ref{fig:fig2} and~\ref{fig:fig3} the quantity $\Delta{E}$/$|E|$ corresponds to the angle(s) of the polygon in the final state (i.e., the lower right panels) with dashed sides.  We caution that at high virial ratios, very 
few triples form, and small-number statistics become a concern in the green 
histograms.  Note that the distributions of single star escape velocities are noticeably different for encounters that produce stable triples, 
relative to those that produce two single stars and a binary.  

\begin{figure*}
\begin{center}                                                                                                                                                           
\includegraphics[width=\textwidth]{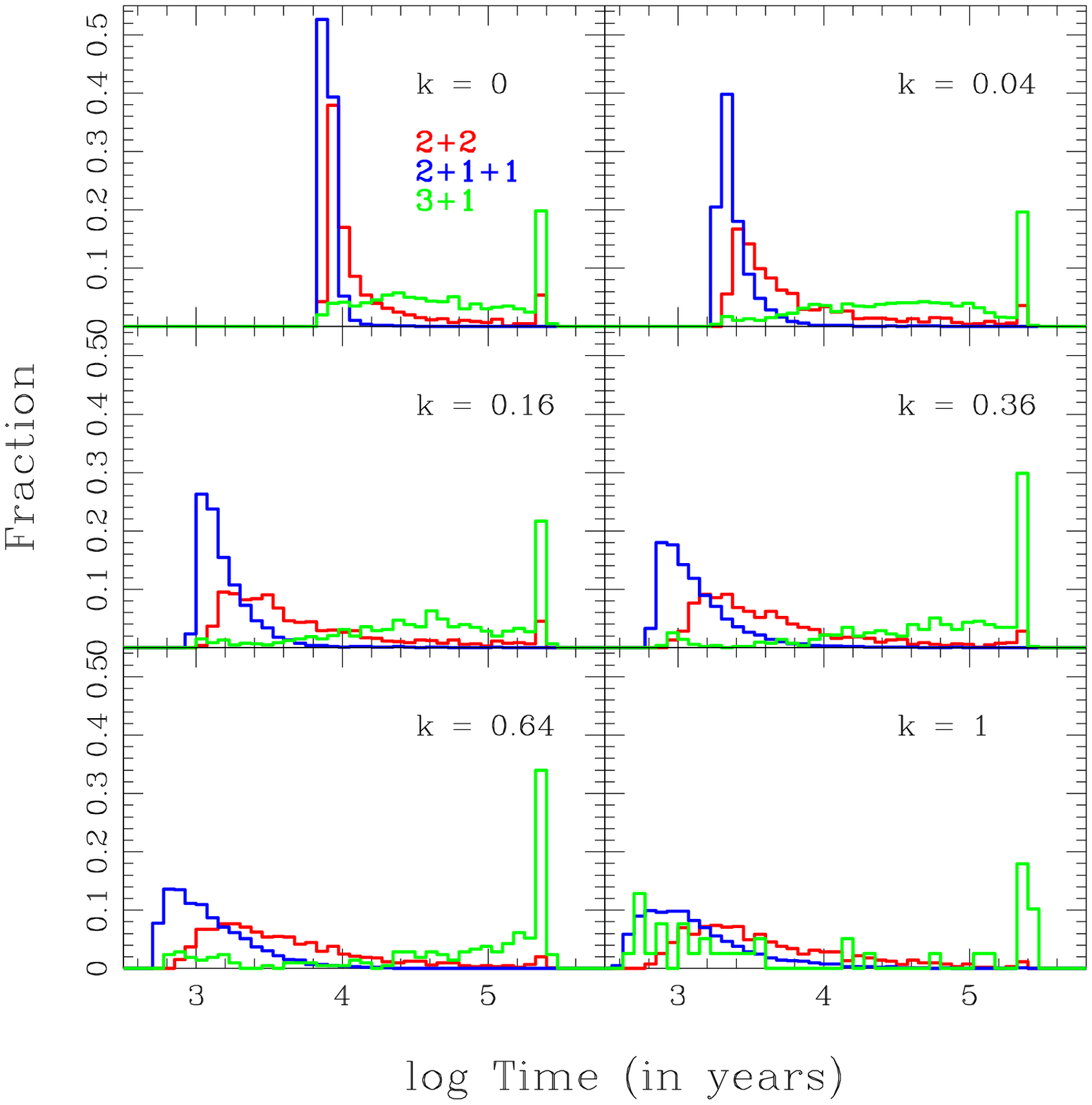}
\end{center}
\caption[Distributions of total encounter durations for every initial virial ratio]{The distributions of total encounter durations 
are shown in years.  The blue, red 
and green histograms correspond to the 2+1+1, 2+2 and 3+1 outcomes, respectively.  Each panel shows the 
distributions for a different value of the initial virial ratio.  All histograms have been normalized by the total number of simulations that resulted in the corresponding outcome.
\label{fig:fig8}}
\end{figure*}

In Figure~\ref{fig:fig8}, we show the distributions of total encounter 
durations (in years) for each of our three encounter outcomes.  Note that the initial displacement toward longer encounter durations, which is more severe for low virial ratios, is due to the initial infall time for the two binaries, which is determined by our choice for the tidal tolerance parameter (and is hence not physical).  The encounter durations are shortest for the the 2+1+1 outcome, and 
longest for the 3+1 outcome.  This is at least due in part to the fact that it is the outer orbit of the top-level node in \texttt{FEWBODY} that decides the final stability of the system, and the encounter cannot be completed until one orbital period has occurred.  Notwithstanding, these results serve to further strengthen the conclusion that each outcome can 
be regarded, to first order, as a unique transformation in energy and angular momentum space, since each outcome produces 
distinct distributions of virial ratios, escape velocities and encounter durations.

\begin{figure*}
\begin{center}                                                                                                                                                           
\includegraphics[width=\textwidth]{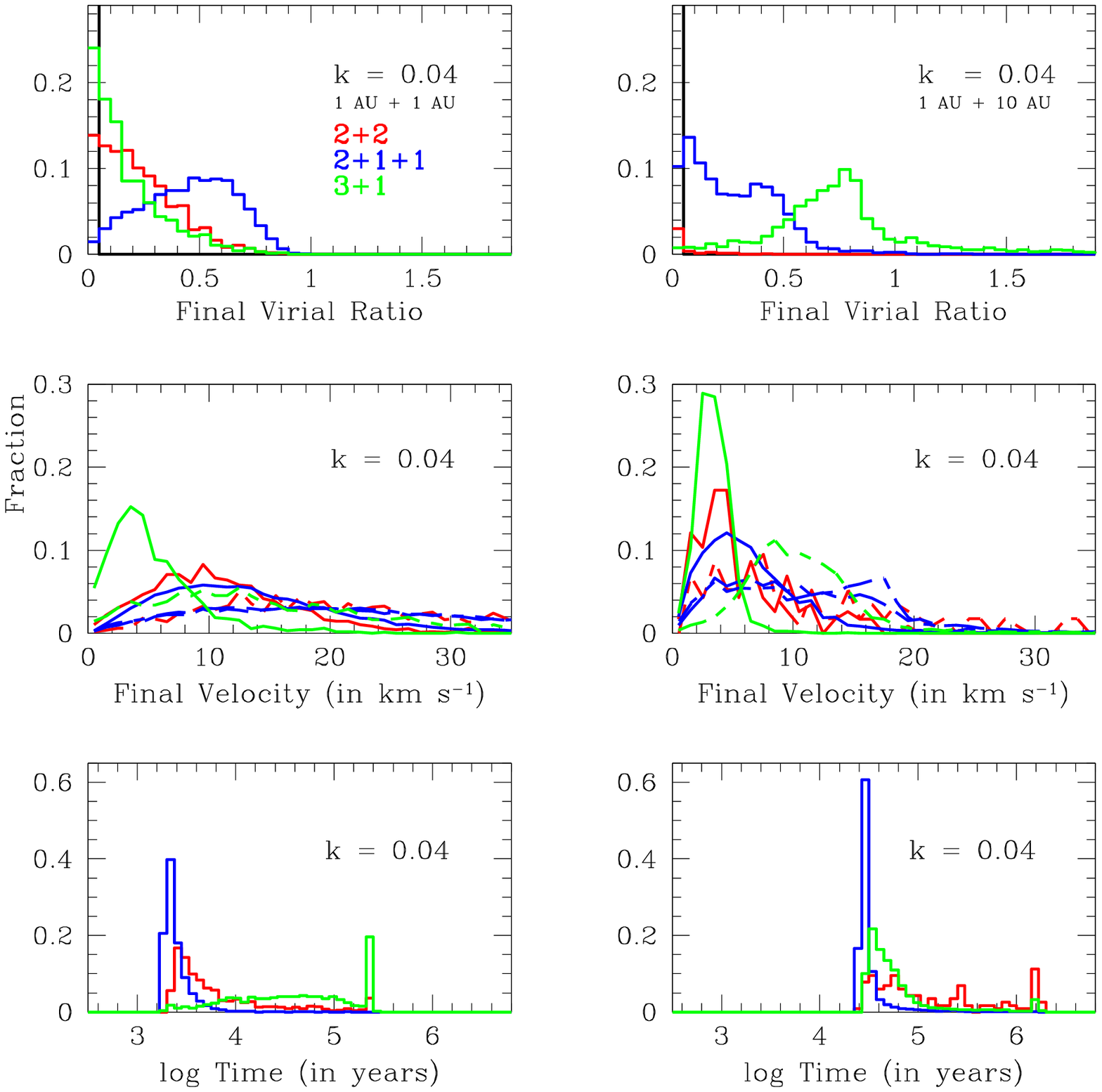}
\end{center}
\caption[Distributions of final virial ratios, escape velocities and total encounter durations for $k = 0.04$]{The distributions of final virial ratios (top panels), escape velocities (in km s$^{-1}$; middle panels) and total encounter durations (in years; bottom panels) for an initial virial ratio of $k = 0.04$.  The left panels show the results for simulations involving two binaries with identical initial separations of $a_{\rm 1} = a_{\rm 2} =$1 AU, whereas the right panels show the results for simulations involving binaries with initial separations of $a_{\rm 1} =$ 1 AU and $a_{\rm 2} =$10 AU.  The blue, red 
and green histograms correspond to the 2+1+1, 2+2 and 3+1 outcomes, respectively.  All histograms have been normalized by the total number of simulations that resulted in the corresponding outcome.
\label{fig:fig9}}
\end{figure*}

Finally, in Figure~\ref{fig:fig9}, we compare our fiducial simulations involving two identical binaries with initial separations of $a_{\rm 1} = a_{\rm 2} =$ 1 AU (left panels) to what we obtain assuming instead that the binaries have initial separations of $a_{\rm 1} =$ 1 AU and $a_{\rm 2} =$10 AU (right panels).  This is done for one choice of the initial virial ratio, 
namely $k = 0.04$, since the computational expense for these additional simulations is even higher than for our fiducial runs.  In the top, middle and bottom insets we show, respectively, the distributions of final virial ratios, escape velocities (in km s$^{-1}$) and total encounter durations (in years).  Where comparisons are possible, our results are in overall good agreement with those of \citet{mikkola84}, who performed a similar study of encounters involving binaries with unequal energies.  

A few trends are immediately clear from the comparisons shown in Figure~\ref{fig:fig9}.  First, the fraction of simulations producing stable triples is higher by a factor $\gtrsim$ 2.  Correspondingly, the distributions of escaper velocities have been shifted to lower velocities for the 1 AU $+$ 10 AU case which, as previously illustrated, are more conducive to triple formation.  In the bottom panels, we see that the encounter durations corresponding to triple formation are much shorter for the 1 AU $+$ 10 AU case, relative to our fiducial simulations, whereas encounters producing two binaries tend to have longer encounter durations.  We speculate that this is due to the higher angular momentum characteristic of these additional simulations with 1 AU $+$ 10 AU binaries; in order to conserve angular momentum any ejected binary will be most likely to have a large orbital separation and hence a small absolute orbital energy.  This puts it very close to the dissociation border, such that even a small additional amount of positive energy will dissociate the binary, which likely results from tidal effects imparted by the remaining (more compact) binary.  More work needs to be done to better understand the underlying physics responsible for the differences highlighted in Figure~\ref{fig:fig9}, which arise due to the different ratios between the initial binary orbital separations.  The take-away message from this figure is that larger ratios between the initial binary orbital separations correspond to a higher probability of triple formation at a given virial ratio, relative to what is shown for our fiducial simulations.    

\section{Model} \label{model}


\subsection{Time-averaged virial approximation} \label{virial}

Consider a small star cluster of $N$ identical point-particles each with mass $m$.  If the system is in dynamical equilibrium, 
then the mean 
radius $R$ of the cluster is determined by the virial radius \citep{valtonen06}:
\begin{equation}
\label{eqn:virialR}
R \sim \frac{GM}{v_{\rm rms}^2} = \frac{GM^2}{|E|},
\end{equation}
where $M = Nm$ is the total cluster mass, $v_{\rm rms} =$ ($|E|/M$)$^{1/2}$ is the the root-mean-square velocity of the virialized system and E is its total  energy.  Let us assume that a single star leaves the cluster, escaping to spatial infinity with a positive total energy ${\Delta}E$.  After the system re-achieves dynamical equilibrium, the new virial radius $R'$ is:
\begin{equation}
\label{eqn:virialR2}
R' \sim \frac{G(M-{\Delta}M)^2}{|E|+{\Delta}E},
\end{equation}
where ${\Delta}M >$ 0 and $\Delta{E}$ is positive if the escaping object is a single star, or if the escaping object is a binary and the absolute value of its orbital energy is less than its (translational) kinetic energy.  Equation~\ref{eqn:virialR2} can be re-written:
\begin{equation}
\begin{gathered}
\label{eqn:virialR3}
R' = \frac{GM^2}{E}\frac{(1 - {\Delta}M/M)^2}{(1 + {\Delta}E/|E|)} \\
     = R\frac{(1 - {\Delta}M/M)^2}{(1 + {\Delta}E/|E|)} \\
     = {\alpha}R.
\end{gathered}
\end{equation}
If $\alpha >$ 1 ($< 1$), then the left-over system expands (contracts).  Note that $\alpha > 1$ (and hence expansion) requires 
${\Delta}E/|E| <$ 0, and hence can only occur when a binary is the escaping object (i.e., for a 2+2 outcome).  In encounters featuring very close pericenter passages, orbital expansion can also in principle occur through dissipative processes such as gravitational wave emission \citep{peters64} or tidal excitation of stellar modes \citep{press77}, but we neglect finite-size and general relativistic effects in this work.  In general, from Equation~\ref{eqn:virialR3}, we 
see that for ejections corresponding to very high escape velocities but low particle masses compared to $M = Nm$ 
(i.e., $|{\Delta}E/E|$ $\gg$ $|{\Delta}M/M|$) the system should expand.  Conversely, if 
$|{\Delta}E/E|$ $\sim$ $|{\Delta}M/M|$, the system will typically contract.   

Returning to the four-body problem, if ${\Delta}E >$ 0, which is always the case for the escape of a single 
star (since we have taken 
the absolute value of the system binding energy $|E|$ in Equation~\ref{eqn:virialR3}), then $\alpha <$ 1 
and the system contracts.  But, contraction is not ideal for promoting stable triple formation, since it decreases the 
probability of forming a significant hierarchy with a stable outer single star orbiting a compact inner binary.  Thus, 
we might naively expect from this that, when the first escape of a single star occurs, encounters with the lowest 
escape velocities and hence the lowest fractional change in energy should be more likely to 
produce stable triples.  This is exactly what is seen in Figure~\ref{fig:fig7}.  
That is, most stable triples are formed 
when a single star escapes with a very low $\Delta{E}$/$|E|$ value.  However, the converse is not always true, 
due to angular momentum conservation and the chaotic nature of the system evolution.  In other words, if a single 
star escapes with a low $\Delta{E}$/$|E|$ value, it is not guaranteed that a stable triple will form.  As shown 
in Figure~\ref{fig:fig7}, at low virial 
ratios, stable triples are only more likely to be associated with low $\Delta{E}$/$|E|$ values by a factor of 
$\sim$ a few, whereas this factor increases markedly with increasing virial ratio.  Finally, as shown in 
Figure~\ref{fig:fig7}, due to conservation of energy and linear momentum, the vast majority of encounters leading to 
stable triple formation correspond to escape events satisfying $\Delta{E}$/$|E| \ll$ 1.  

Conversely, if the first escape of a single 
star leaves behind an unstable triple, then a subsequent escape event will occur, leaving behind 
two single stars and a binary.  These events should preferably correspond to the highest velocities 
for the escape of the first particle, and hence the highest values of $\Delta{E}$/$|E|$ (as shown in Figure~\ref{fig:fig7}) 
since these tend to result in the greatest contraction of the system post-escape (but not always; see Figure~\ref{fig:fig7}).  
More compact three-body configurations have a lower probability of being stable, due to the smaller parameter 
space corresponding to stability that is available to them.

\subsection{Application of the Monaghan formalism to the 4-body problem} \label{monaghan}

Next, we adapt the Monaghan formalism to the four-body problem.  This was originally developed by \citet{monaghan76a} 
for the three-body problem to estimate the statistical distribution of outcomes from three-body scatterings.  We make a historical note here for completeness, namely that the Monaghan distribution derived a power-law index of -5/2 for the distribution of binary binding energies (originally derived by \citet{jeans28}).  This was shown to provide a poor fit to numerical data, but was later corrected by \citet{valtonen06} using the loss-cone approximation, changing the power-law index to -9/2 (originally derived by \citet{heggie75} and confirmed in \citet{saslaw74}).

The Monaghan formalism relies on the density of escape configurations per unit energy 
$\sigma$, obtained by integrating over the phase space volume.  As we will show, this can be done relatively precisely for the 
3+1 outcome (i.e. the escaping object is a single star and the left-over system is a stable hierarchical triple), whereas this is not the 
case for the other two encounter outcomes.  This is due to a strong parallel that can be drawn between the 3+1 outcome and the 
escape of a single star during a three-body interaction, as the outer orbit of the final stable triple contains a negligible 
total energy.  We will return to this point in the subsequent sub-section.

Given that there are three possible outcomes for the $E < 0$ four-body problem (with negative total energy), we write:
\begin{equation}
\label{eqn:Punity}
1 = P_{\rm 2+2} + P_{\rm 3+1} + P_{\rm 2+1+1},
\end{equation}
where $P_{\rm 2+2}$, $P_{\rm 3+1}$ and $P_{\rm 2+1+1}$ are the probabilities corresponding to the encounter outcomes, respectively, 2+2, 3+1 and 2+1+1.  Equation~\ref{eqn:Punity} can also be re-written as:
\begin{equation}
\label{eqn:Punity2}
1 = \int \Big( \sigma_{\rm 2+2}(|E|) + \sigma_{\rm 3+1}(|E|) + \sigma_{\rm 2+1+1}(|E|)\Big)d|E|,
\end{equation}
where each $\sigma$ value denotes the (normalized) density of escape configurations per unit energy for each of the indicated outcomes.  We will return to Equation~\ref{eqn:Punity2} in Section~\ref{monaghan}.

\subsubsection{Deriving the distribution of ejection velocities} \label{eject}

Consider a chaotic four-body interaction that produces a stable hierarchical triple system.  The total energy 
of the four-body system is:
\begin{equation}
\label{eqn:eqn1}
E_{\rm 0} = E_{\rm e} + E_{\rm b},
\end{equation}
where the total encounter energy is divided between the escaper energy $E_{\rm e}$, and 
the energy of the left-over (bound) system $E_{\rm b}$ which is in this case a stable triple.\footnote{Note that we 
have replaced the variables $E_{\rm s}$ and $E_{\rm B}$ in the original Monaghan derivation with, respectively, 
$E_{\rm e}$ and $E_{\rm b}$.}  Hence, the final 
system energy can be decomposed into the inner and outer orbital energies:
\begin{equation}
\label{eqn:eqn2}
E_{\rm b} \approx E_{\rm b,a} + E_{\rm b,b} \sim E_{\rm b,b},
\end{equation}
where the total system energy can be decomposed into the inner and outer orbital energies:
\begin{equation}
\label{eqn:eqn2}
E_{\rm b,a} = -\frac{Gm_{\rm a,1}m_{\rm a,2}}{2a_{\rm b,a}}
\end{equation}
and
\begin{equation}
\label{eqn:eqn3}
E_{\rm b,b} = -\frac{Gm_{\rm b,1}m_{\rm b,2}}{2a_{\rm b,b}},
\end{equation}
where $m_{\rm b} =$ $m_{\rm b,1} +$ $m_{\rm b,2}$.  We assume that $a_{\rm b,a} =$ ${\beta}a_{\rm b,b}$ for 
some positive constant $\beta \gtrsim$ 10 (i.e. for 
a stable hierarchical triple; \citealt{mardling01}).  This gives, to first order, $E_{\rm b} \sim$ $E_{\rm b,b}$.  
For now, we stick with the variable $E_{\rm b}$ for the rest of the derivation of the distribution of single star escape velocities, 
to highlight the analogy with the three-body problem. 

The density of escape configurations per unit energy is:
\begin{equation}
\label{eqn:sigma}
\sigma_{\rm 3+1} = {\int} {\rm ...} {\int} {\delta}\Big(\frac{p_{\rm e}^2}{2m} + V_{\rm e} + E_{\rm b} - E_{\rm 0} \Big){d\bf{r_{\rm e}}}{d\bf{p_{\rm e}}}{d\bf{r}}{d\bf{p}}, 
\end{equation}
where 
\begin{equation}
\label{eqn:eqn4}
m = \frac{m_{\rm b}m_{\rm e}}{M} 
\end{equation}
and 
\begin{equation}
\label{eqn:eqn5}
M = m_{\rm b}+m_{\rm e}
\end{equation}
Now, assuming (49/4)($a_{\rm b,a}$/$r_{\rm e}$)$^2$ for the loss-cone factor as in \citet{valtonen06},\footnote{We note that this loss-cone factor was determined from 1+2 scattering experiments, and should be tested specifically for the four-body problem in future studies concerned with the normalization factor in front of the integral in Equation~\ref{eqn:sigma4}.  This normalization factor enters into the "branching ratio", or the probability of the encounter ending in a given outcome.  Crucially, our assumption for this loss-cone factor does not affect our derived velocity distributions, only the normalization factor or "branching ratio".} the integrations over $\bf{p_{\rm e}}$ 
and $\bf{r_{\rm e}}$ are carried out analogously and we obtain:
\begin{equation}
\label{eqn:sigma2}
\sigma_{\rm 3+1} = 98\sqrt{2}{\pi}^2(GMR)^{1/2}m^2(Gm_{\rm b,1}m_{\rm b,2})^2{\int} {\rm ...} {\int}\frac{{d\bf{r}}{d\bf{p}}}{|E_{\rm b}|^2},
\end{equation} 
where the upper limit $R$ of the integration over the $r_{\rm e}$ parameter is a free parameter, but must be chosen 
to be relatively small (i.e., taken to be $R =$ 3$a_{\rm 0}$ in \citealt{valtonen06}, where $a_{\rm 0}$ is the initial semi-major axis of the binaries going into the four-body interaction, which we assume for simplicity are equal).  

Here, we recall the approximation $E_{\rm b} \sim$ $E_{\rm b,b}$, which implies that the contribution to the total energy from the outer 
orbit of the stable hierarchical triple can be ignored in the derivation for $\sigma_{\rm 1+3}$.  In other words, all of the total encounter energy is distributed among the escaping star and the inner binary of the triple in the final (stable) configuration.  Only the angular momentum of the outer orbit is significant.  Hence, as we show in the next section, we can model a 3+1 outcome as a single escape event, with $m_{\rm b} =$ $m_{\rm b,b} =$ $m_{\rm b,1} +$ $m_{\rm b,2}$ and $m_{\rm e} =$ $M - m_{\rm b}$, or 
$m_{\rm e} =$ $m_{\rm b} =$ $M$/2 for all identical particles.  In other words, we model a 3+1 outcome using the same formulation as originally developed by \citet{monaghan76a} for the three-body problem, but modified slightly by assuming that two stars are effectively "ejected" together.  One of these stars escapes to infinity and the other ends up loosely bound to the remaining binary.  Only the escaping single star and the inner binary of the triple exchange significant energy, while the remaining star becomes the outer triple companion to the binary and carries only a negligible amount of the total encounter energy.

The rest of the derivation for the 3+1 outcome is effectively identical to the original Monaghan formalism, given our assumption that only the inner binary of the triple and the escaping single star exchange significant energy.  Hence, we do not repeat the full derivation here, and only describe the essential steps (see \citealt{valtonen06} for more details).  Adopting spherical polar coordinates (r, $\theta$, $\Phi$), the (inner) binary energy can then be written:
\begin{equation}
\label{eqn:Eb}
E_{\rm b} = \frac{1}{2}\frac{p^2}{\mu}-\frac{Gm_{\rm b,1}m_{\rm b,2}}{r},
\end{equation}
where $\mu =$ $m_{\rm b,1}$$m_{\rm b,2}$/($m_{\rm b,1}+m_{\rm b,2}$).  Following \citet{valtonen06}, several integrations can be carried out, leaving only the integrals over the binary energy $E_{\rm b}$ and angular momentum L, or:
\begin{equation}
\label{eqn:sigma3}
\sigma_{\rm 3+1} = 2 \times 98{\pi}^5(Gm_{\rm b,1}m_{\rm b,2})^{7/2}R^{1/2}m_{\rm b}^{3/2}M^{-3/2}m_{\rm e}^2{\int}{\int}\frac{{d}|E_{\rm b}|}{|E_{\rm b}|^{7/2}}L{d}L.
\end{equation}
Re-writing Equation~\ref{eqn:sigma3} in terms of the binary eccentricity e gives, finally:
\begin{equation}
\label{eqn:sigma4}
\sigma_{\rm 3+1} = 98{\pi}^5(Gm_{\rm b,1}m_{\rm b,2})^{11/2}R^{1/2}m_{\rm b}^{3/2}M^{-3/2}m_{\rm e}^2{\mu}{\int}{\int}\frac{{d}E_{\rm b}}{|E_{\rm b}|^{9/2}}e{d}e.
\end{equation}

The quantities following the integral signs in Equation~\ref{eqn:sigma4} correspond to the distributions over which one must 
integrate in order to 
obtain the total phase space volume corresponding to the 3+1 outcome.  Thus, the distribution of binary energies $|E_{\rm b}|$, 
normalized to unity, is:
\begin{equation}
\label{eqn:distEb}
f(|E_{\rm b}|){d}|E_{\rm b}| = 3.5|E_{\rm 0}|^{7/2}|E_{\rm b}|^{-9/2}{d}|E_{\rm b}|.
\end{equation}
This corresponds to the low total angular momentum case.  More generally, numerical scattering experiments have shown 
that, in order to cover the full range in total angular momenta, Equation~\ref{eqn:distEb} can be re-written \citep{valtonen06}:
\begin{equation}
\label{eqn:distEb2}
f(|E_{\rm b}|){d}|E_{\rm b}| = (n-1)|E_{\rm 0}|^{n-1}|E_{\rm b}|^{-n}{d}|E_{\rm b}|,
\end{equation}
where, for the three-body problem, the power-law index $n$ ranges from $n =$ 3 at $L_{\rm 0} =$ 0 to $n =$ 14.5 
at $L_{\rm 0} =$ 0.8$L_{\rm max}$, and $L_{\rm 0}$ is the total encounter angular momentum \citep{valtonen06}.  Hence, 
$n$ is a positive constant that depends on the total angular momentum via the substitution $L = L_{\rm 0}$/$L_{\rm max}$, or:
\begin{equation}
\label{eqn:nl}
n - 3 = 18L^2,
\end{equation}
as found from numerical scattering experiments for the three-body problem \citep{valtonen03}.  Importantly, we have 
not verified in this paper that Equation~\ref{eqn:nl} holds for the chaotic four-body problem, and this should be tested 
in future work.  To do this, significantly more simulations must be performed, in order to sample the full range of angular 
momentum space.  In this paper, we focus on the minimum angular momentum case (i.e., zero impact parameter and $n =$ 3 
in Equation~\ref{eqn:nl}), for which Equation~\ref{eqn:nl} does indeed provide a good match to the simulations (see 
Figures~\ref{fig:fig10} and~\ref{fig:fig11}).

Equation~\ref{eqn:distEb2} can ultimately be used to derive the distribution of escape velocities for the escaping single 
star produced during a 3+1 outcome, as given by Equation~\ref{eqn:vdist} with $m_{\rm e} =$ $m_{\rm b} =$ $M$/2 for all 
identical particles.  That is, the integration in Equation~\ref{eqn:sigma4} can 
instead be re-written to be carried out over the single star escape velocity $v_{\rm e}$ to obtain the escape velocity distribution 
$f(v_{\rm e})dv_{\rm e}$.  This gives the following functional form (Equations 7.19 and 7.26 in \citealt{valtonen06}):
\begin{equation}
\label{eqn:vdist}
f(v_{\rm e})dv_{\rm e} = \frac{(n|E_{\rm 0}|^{n-1}(m_{\rm e}M/m_{\rm b}))v_{\rm e}dv_{\rm e}}{(|E_{\rm 0}| + \frac{1}{2}(m_{\rm e}M/m_{\rm b})v_{\rm e}^2)^{n}}.
\end{equation}


As we will show, the above derivation works only to reproduce the distribution of escape velocities for 3+1 outcomes.  For the 2+2 and 2+1+1 outcomes, the escape velocity distributions are reproduced by Equation~\ref{eqn:vdist} at low virial ratios only.  This is because 
the assumption $E_{\rm b} \sim E_{\rm b,b}$ can only be applied for the 3+1 outcome.  In the 2+2 and 2+1+1 cases, this 
approximation is not valid, 
since significant additional energy can be stored in either of the two remaining orbits (after the first ejection event, whether it be a single 
star or a binary).  This seems to account for the additional positive energy imparted at large virial ratios to the escaping object in the 2+2 
and 2+1+1 outcomes, relative to the Monaghan formalism (see Figures~\ref{fig:fig9} and~\ref{fig:fig10}).   We will return to this 
important and interesting result in Section~\ref{discussion}.  For now, we note that we intend to apply this theory to the other two encounter outcomes in a forthcoming paper.  This will ultimately require integrating over the additional degrees of freedom brought in by the additional binary orbit (i.e., relative to the more familiar three-body problem).

\subsubsection{Fitting to the simulated data} \label{fitting}

\begin{figure*}
\begin{center}                                                                                                                                                           
\includegraphics[width=\textwidth]{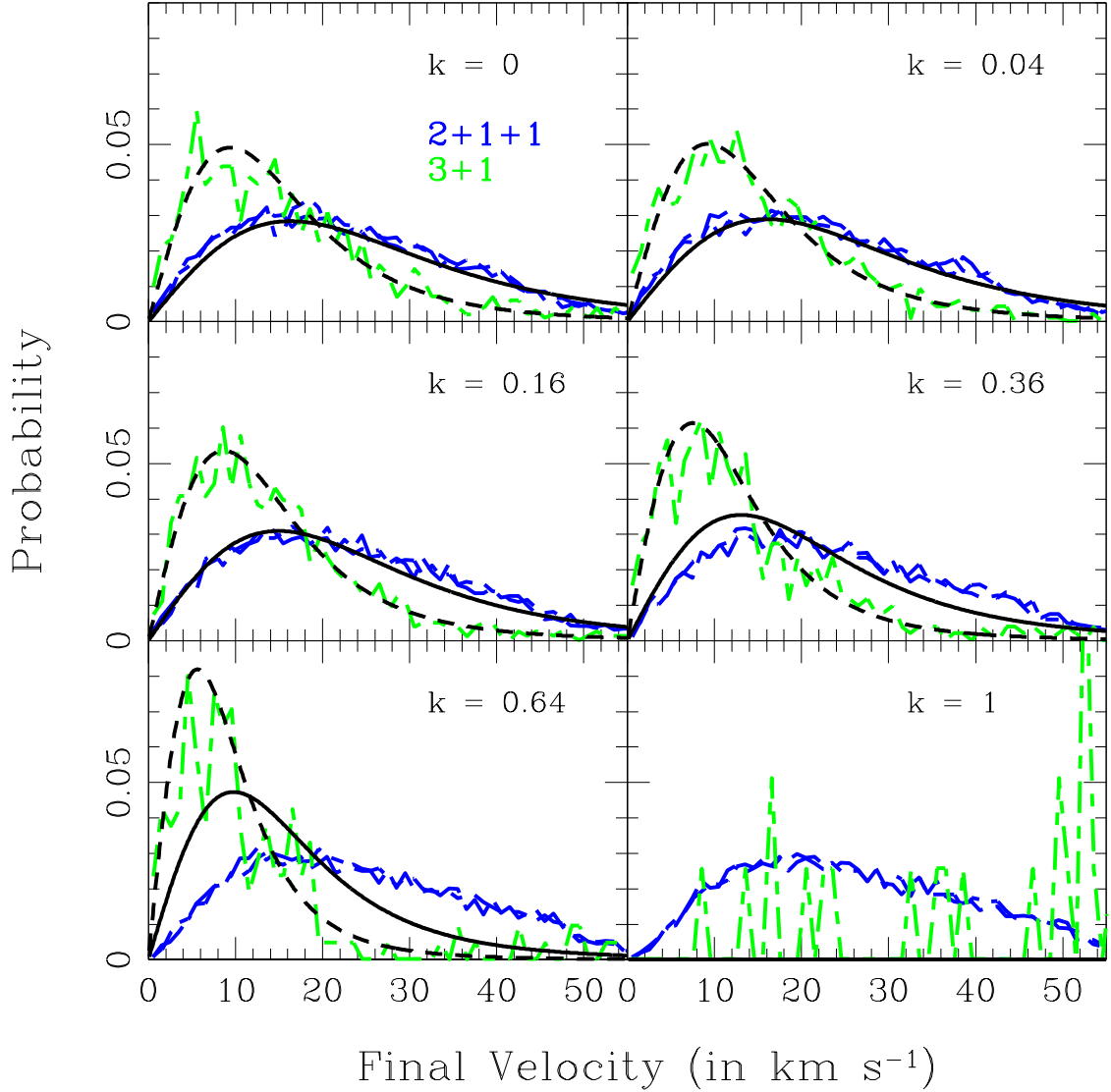}
\end{center}
\caption[Distributions of single star escape velocities for every initial virial ratio for both the 3+1 and 2+1+1 outcomes]{The 
distributions of single star escape velocities are shown in km s$^{-1}$ for the 3+1 (green) and 2+1+1 (blue) outcomes.  The 
black lines show the distribution of escape velocities calculated using Equation~\ref{eqn:vdist} for a 3+1 outcome and 
assuming $n =$ 3 (which corresponds to the minimum angular momentum case in Equation~\ref{eqn:nl}).  For the solid and 
dashed black lines we assume, respectively, $m_{\rm e} =$ $m_{\rm b}$/3 $=$ $M$/4 and 
$m_{\rm e} = m_{\rm b}$ $=$ $M$/2.  The different insets show the distributions for different 
virial ratios, as indicated.
\label{fig:fig10}}
\end{figure*}

\begin{figure*}
\begin{center}                                                                                                                                                           
\includegraphics[width=\textwidth]{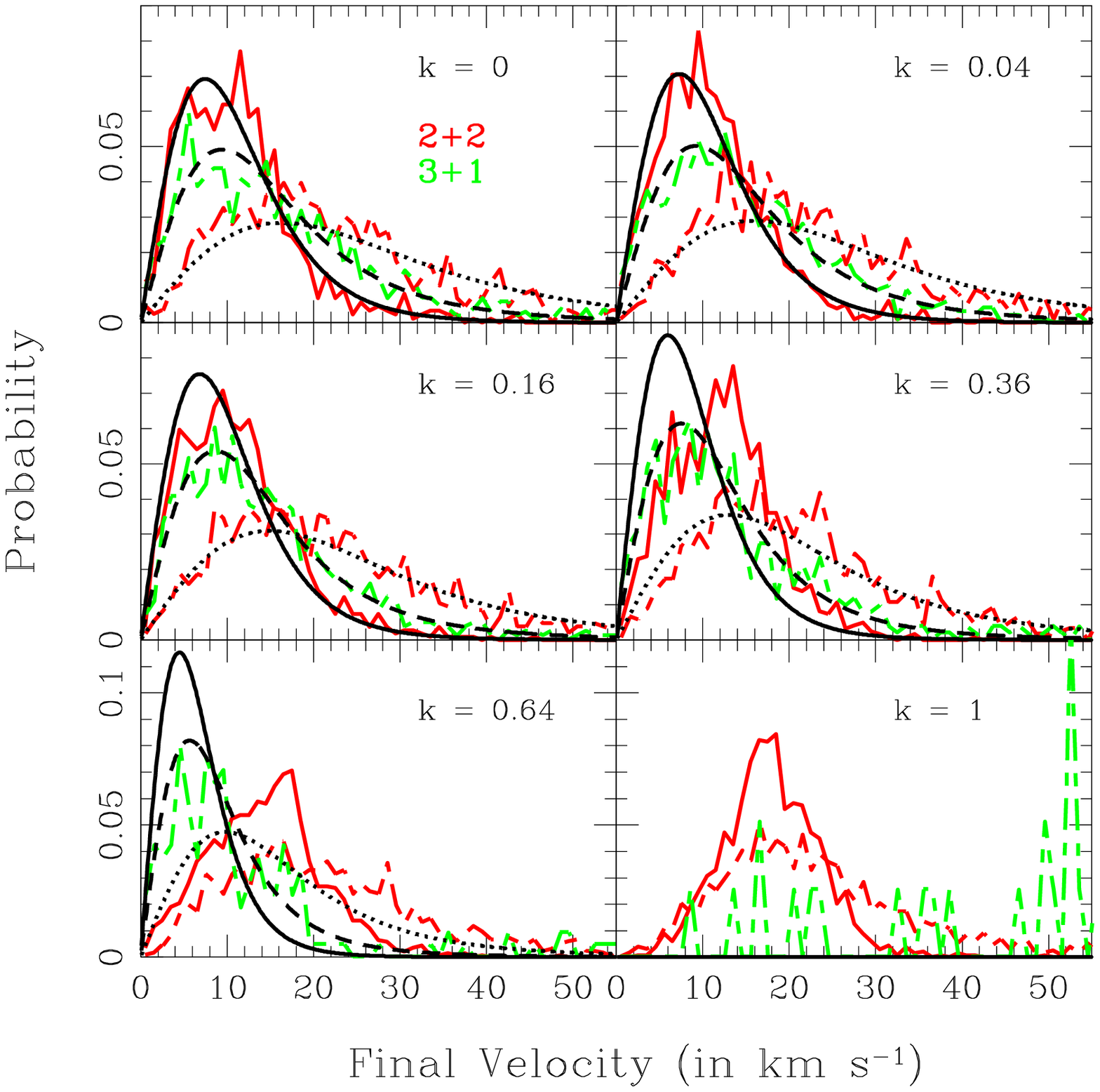}
\caption[Distributions of binary star escape velocities for every initial virial ratio for the 2+2 outcome]{The 
distributions of binary star escape velocities are shown in km s$^{-1}$ for the 2+2 outcomes are shown by 
the solid and dashed red lines.  The solid black lines show the distribution of escape velocities calculated using 
Equation~\ref{eqn:vdist} and assuming $n =$ 4.5 (this corresponds to small but non-zero total angular momentum, via 
Equation~\ref{eqn:nl}).  For comparison, we also show the single star escape velocity distributions for a 3+1 
outcome and assuming as before $n =$ 3, with both $m_{\rm e} =$ $m_{\rm b}$/3 $=$ $M$/4 
(dashed black lines) and $m_{\rm e} =$ $m_{\rm b}$ $=$ $M$/2 (dotted black lines).  
The different insets show the distributions for different 
virial ratios, as indicated.
\label{fig:fig11}}
\end{center}
\end{figure*}

In this section, we compare the theoretical distributions of escaper velocities derived in the previous section for the 3+1 
outcome to the results of our numerical scattering experiments.  For comparison, we also show the distributions of 
escaper velocities for the 2+1+1 (see Figure~\ref{fig:fig10}) and 2+2 (see Figure~\ref{fig:fig11}) outcomes, beginning 
with the former.

In Figure~\ref{fig:fig10} we show, for different initial virial ratios, the distribution of escape velocities for the 
single star for a 3+1 outcome (green lines), as well as for both single stars for a 2+1+1 outcome (solid and dashed 
blue lines).  The key point 
to take away from Figure~\ref{fig:fig10} is that escaping single stars that leave behind a stable triple have a unique 
velocity distribution relative to the corresponding distributions for the escaping single stars leaving behind a binary.  
Note that the increased scatter in the green histograms seen at large virial ratios is due to the fact that the number 
of simulations resulting in a 3+1 outcome is very small here (i.e., $\sigma_{\rm 3+1} \rightarrow$ 0 as 
$E_{\rm 0} \rightarrow$ 0).

The black lines in Figure~\ref{fig:fig10} show our best fits to the distributions of escaping single star velocities for the 3+1 and 
2+1+1 outcomes, given by Equation~\ref{eqn:vdist}.  As shown in Figure~\ref{fig:fig10} by the dashed black line, for the 3+1 
outcome we find a good match between the theoretical and simulated distributions for all virial ratios.  Hence, the original Monaghan 
formalism, as derived for the three-body problem, is directly applicable to the formation of stable triples (i.e., the 3+1 
outcome) during the chaotic four-body interaction, provided we assume that both the escaping single star and the outer third 
companion of the resulting triple are effectively "ejected" together, with only one star obtaining a final velocity $v_{\rm e} >$ 0 
(in order to escape the system) and the other ending with a final velocity slightly less than the system escape speed (such that its 
final orbital energy is nearly zero).  

The peak of the distribution in Equation~\ref{eqn:vdist} can be found by setting $df/dv_{\rm e} =$ 0.  The 
result can be written in the general form:
\begin{equation}
\label{eqn:vpeak}
v_{\rm e,peak} = \epsilon\sqrt{\frac{(M-m_{\rm e})}{m_{\rm e}M}}\sqrt{|E_{\rm 0}|},
\end{equation}
where $\epsilon$ is a positive constant of order unity equal to:
\begin{equation}
\label{eqn:epsilon}
\epsilon = \frac{1}{\sqrt{n-\frac{1}{2}}}.
\end{equation}  

Equation~\ref{eqn:vdist} provides a reasonable fit to the simulated escape velocity distributions for the 2+1+1 outcome 
only at low virial ratios, as shown by the solid black lines in Figure~\ref{fig:fig10}.  The agreement 
gets worse as we go to larger virial ratios, and 
begins to better approximate the simulated escape velocity distribution for the 3+1 outcome.  
At the same time, the distribution of single star escape velocities for the 2+1+1 outcome is remarkably insensitive to the initial 
virial ratio (note that the distributions for both escaping single stars overlap and are almost identical for all virial ratios), as 
originally found by \citet{mikkola83}.  This 
is a curious result worthy of further investigation.  As pointed out by \citet{mikkola83}, this trend can perhaps be understood by noting that the mean change in the total binding energy decreases roughly linearly with increasing impact energy, or virial ratio.      

We see a similar scenario in the distributions of binary escape velocities for 
the 2+2 outcome in Figure~\ref{fig:fig11}, as shown by the solid and dashed red lines.  That is, Equation~\ref{eqn:vdist} provides 
a good agreement to the simulated data at low virial ratios only, as shown by the solid and dotted black lines.  At larger 
virial ratios, the agreement becomes significantly 
worse.  Unlike the blue histograms in Figure~\ref{fig:fig10}, however, the red histograms in Figure~\ref{fig:fig11} do change 
with increasing virial ratio, shifting to slightly higher binary escape velocities.  Hence, at large virial ratios, Equation~\ref{eqn:vdist} 
under-predicts the peak escaper velocity.

Only the distribution of single star 
escape velocities for the 3+1 outcome, shown by the green histograms, matches Equation~\ref{eqn:vdist} for all virial 
ratios.  This is illustrated via the solid and dashed black lines in, respectively, Figures~\ref{fig:fig10} 
and~\ref{fig:fig11}.  We will return to this intriguing result in Section~\ref{discussion}.  

\section{Discussion} \label{discussion}

The key point to take away from the preceding sections is that the different outcomes of the chaotic four-body problem 
generally correspond to different discrete transitions in energy and angular momentum space.  Consequently, the distributions 
of escape velocities are unique for the different encounter outcomes.  Thus, these distributions can potentially 
be used to constrain the origins of dynamically-formed populations, whether they are single, binary or triple stars.  
This is done via a direct comparison between the predicted and observed velocity distributions.  For example, 
comparisons can be made to populations of single stars thought to be connected to triple formation.  These types 
of four-body encounters have also been proposed to produce hypervelocity stars via the capture of binaries around massive 
black holes when they interact with interloping triple stars \citep{perets09}.  Conversely, comparisons can be made to 
the observed velocity 
distributions of triples thought to be formed dynamically, preferably in massive clusters with very long relaxation times. 

Interestingly, the distributions of escape velocities for the chaotic four-body problem are surprisingly similar to 
the distributions originally derived by \citet{monaghan76a} for the three-body problem.  \textit{The single star escape 
velocity distributions during stable triple formation (i.e., 3+1 outcomes) are well described by the Monaghan formalism 
at all virial ratios and arbitrary total angular momentum, provided we assume that both the escaping single star and the 
outer third companion of the resulting triple are effectively "ejected" together}, analogous to a 2+2 outcome.  
The escape velocity distributions for the 2+1+1 and 2+2 outcomes are also well described by the Monaghan formalism, 
but only at low virial ratios 
(i.e., $k \lesssim$ 0.5).  At larger virial ratios, the Monaghan distributions over-predict the fraction of low-velocity 
escapers relative to the simulations. 

Importantly, within the parameter space explored here 
$\gtrsim$ 80\% of all encounters between identical binaries result in the 2+1+1 outcome, as shown in Figure~\ref{fig:fig4}.  As 
shown in Figure~\ref{fig:fig8}, the resulting distribution of single star escape velocities is well described by the 
same escape velocity distribution as originally derived by \citet{monaghan76a} for the three-body problem, 
provided the virial ratio is small.  Thus, 
any population of single and binary stars in a cluster that have gone through \textit{either} a single-binary (i.e., 1+2) 
or binary-binary (i.e., 2+2) encounter should have post-encounter velocities that are roughly described by 
Equation~\ref{eqn:vdist}, assuming that at least one relatively hard binary is included (to ensure a small virial 
ratio) in the encounter and no 
other factors influenced the final escape velocity distribution (e.g., relaxation in a star cluster).  In a dense globular 
cluster, for example, the post-encounter single star velocity distribution should be described by averaging $|E_{\rm 0}|$ in 
Equation~\ref{eqn:vdist} over a Maxwellian velocity distribution.  This offers a potential means of identifying 
evidence for dynamical processing in dense, massive star clusters with very long relaxation times.  That is, 
we hypothesize that if the relaxation time is very long compared to the time-scale for 1+2 and 2+2 encounters to 
occur, then the velocity 
distributions for both the binaries and single stars in the cluster core should deviate from Maxwellian distributions, 
approaching more and more the distributions given by Equation~\ref{eqn:vdist} as time goes on (until some 
steady-state balance is achieved).     

As stated above, the simulated distribution of single star escape velocities for the 2+1+1 outcome agrees well with the 
Monaghan formalism, but only at low virial ratios.  At larger virial ratios (i.e., near unity), the Monaghan formalism 
over-predicts the fraction of low-velocity escapers relative to the simulated data.  Interestingly, the resulting escaper 
velocity distributions in Figure~\ref{fig:fig9} (especially for the 2+1+1 outcome) appear 
roughly insensitive to the total encounter energy $|E_{\rm 0}|$ (or virial ratio), and only the relative probabilities 
for the 3+1 and 2+1+1 outcomes change.  This suggests that the Monaghan formula for the escape velocity distributions 
at $k \lesssim$ 1 (i.e., $|E_{\rm 0}|$ $\sim$ 0) should also fit the simulated distributions for the 3+1 outcome, nearly 
independent of $|E_{\rm 0}|$ (since there is only a weak dependence on virial ratio).  Similarly, the Monaghan formula 
for the escape velocity distributions 
at $k \sim$ 0 should fit the simulated distributions for the 2+1+1 outcome nearly independent of $|E_{\rm 0}|$.  To summarize, 
ignoring the normalization constants (and hence the "branching ratios" for the three outcomes of the four-body problem),
we have in the limit $|E_{\rm 0}| \rightarrow$ 0 (or, equivalently, k $\rightarrow$ 0) that 
$\tilde{\sigma}_{\rm 2+1+1} \sim \tilde{\sigma}_{\rm 2+1}$ and $\tilde{\sigma}_{\rm 2+2} \sim \tilde{\sigma}_{\rm 2+1}$, where $\tilde{\sigma}_{\rm 2+1}$ corresponds to the density of states in the original three-body 
body formulation of Monaghan, divided by the normalization constant out in front of the integral.  This 
needs to be confirmed in future studies that consider other initial conditions than in this paper.


In a forthcoming paper, we will consider the implications of different particle masses for the 
results presented in this paper.  We can nonetheless use Equation~\ref{eqn:vdist} along with the 
results presented here to extrapolate to extreme mass ratios.  For instance, consider a scenario 
where two of the four interacting stars are much less massive than the other two.  In this case, the 
low-mass particles have the highest probability of either escaping or ending up as the outer companion 
to the remaining stable triple (i.e., the two most massive stars constitute the inner binary of the triple).  If 
we take the limit $m_{\rm e} \rightarrow$ 0 in 
Equation~\ref{eqn:vpeak}, the peak escaper velocity shifts to lower values, and the corresponding 
peak escaper kinetic energy ${\Delta}E$ becomes 
(${\Delta}E$)$_{\rm peak} =$ $\frac{1}{2}m_{\rm e}v_{\rm e,peak}^2$ $= \epsilon^2|E_{\rm 0}|$/2.  
In this paper, however, all particles have the same mass.  This yields a peak escaper kinetic energy of 
(${\Delta}E$)$_{\rm peak} =$ 3$\epsilon^2|E_{\rm 0}|$/8, which is larger than we found in the limit 
$m_{\rm e} \rightarrow$ 0.  As shown in Figure~\ref{fig:fig7}, encounters that result in stable triple 
formation prefer smaller values of 
${\Delta}E/E$.  Thus, we predict that four-body encounters involving two very low mass particles 
should have a higher probability of triple formation than found in this paper, since we assume all 
identical particles.  We predict a similar phenomenology for encounters with a wider range of 
particle masses, since the most probable stars to escape always have the lowest masses.  Hence, 
typically, encounters with a range of particle masses should produce outcomes that satisfy 
$m_{\rm e} \ll m_{\rm b} \sim M$.  These predictions, derived from our results, are in good qualitative 
agreement with previous studies \citep[e.g.][]{saslaw74,mikkola90,valtonen94}.

A similar argument can perhaps be applied to encounters with higher total angular 
momentum since, as shown in Figure 7.11 of \citet{valtonen06} for the three-body problem, a higher 
angular momentum translates into a lower peak escape velocity (which is preferred during triple formation).  
This should be confirmed in future studies.  Likewise, we have already seen that the probability of stable triple 
formation is enhanced when considering encounters of binaries with varying internal energies (see 
Figure~\ref{fig:fig4}).

Putting this all together, the probability 
of forming stable triples during chaotic 4-body encounters should increase relative to the results shown in this paper, 
which focused primarily on equal mass stars in identical binaries with zero angular momentum.  Varying any of these 
idealized assumptions (mass ratio, binary binding energy, total angular momentum) seems likely to increase the 
probability of the 3+1 outcome.  This is of great interest for formation of exotic stars or transient phenomena that can 
be catalyzed through the Kozai mechanism in hierarchical triples; examples include blue stragglers \citep{perets09b}, 
direct-collision Type Ia supernovae \citep{thompson11,kushnir13}, or gravitational waves from compact object inspirals, 
particularly eccentric compact mergers \citep{antonini16}, which may be impossible to produce at interesting rates 
through non-Kozai channels.  Although detailed rate calculations are beyond the scope of this paper, the greater 
understanding of dynamical triple formation rates developed here is a necessary step towards better quantifying 
the production of interesting Kozai-induced exotica in dense stellar systems.

\section{Summary} \label{summary}


In this paper, we study the chaotic four-body problem in Newtonian gravity, assuming point particles and total 
encounter energies $\le$ 0.  The problem has three possible outcomes, and we describe each outcome as a 
series of discrete transformations in 
energy space using the diagrams first presented in Leigh \& Geller (2012; see 
the Appendix).  We further adapt the original Monaghan formalism for the three-body problem to treat 
the chaotic four-body problem, based on the density of escape configurations per unit energy.  This gives 
theoretical predictions for the probability of an encounter producing a given set of outcome parameters.  We focus 
on encounters that produce stable triples for this analytic derivation, since the outer orbit of the triple carries 
negligible total energy.  Hence, for this outcome alone, the addition of a fourth particle to the original Monaghan 
formalism does not add an additional degree of freedom. 

We simulate a series of binary-binary encounters with identical point particles using the \texttt{FEWBODY} code.  
We compare the resulting single star velocity distributions with our analytic 
predictions, and show that each of the three 
encounter outcomes produces a unique escape velocity distribution.  Thus, these distributions can potentially 
be used to constrain the origins of dynamically-formed populations, via a direct comparison between the predicted 
and observed velocity (and binary binding energy) distributions.  Finally, we show that, for encounters 
that form stable triples, the simulated single star escape velocity distributions are well reproduced by our theoretical 
predictions, in the low angular momentum regime.  Interestingly, this is the case for the other two encounter outcomes 
as well, but only at low virial ratios; as encounters become more energetic, the analytic three-body formalism of 
Monaghan breaks down.  This predicts that single (and hard binary) stars processed via single-binary and binary-binary 
encounters in 
dense star clusters should have a unique and computable velocity distribution (that can be calculated) relative to the 
underlying Maxwellian distribution, provided the relaxation time is sufficiently long.

\appendix

\section{The rarity of complete ionizations during 2+2 encounters in star clusters} \label{appendix}

If the total energy of the binary-binary encounter is positive, a fourth outcome becomes possible: a $1+1+1+1$ total dissociation of the two scattered binaries.  However, this outcome is inherently unlikely in any old, collisional stellar system, because all binaries on the soft side of the hard-soft boundary will have long ago dissociated, and the surviving systems that engage in binary-binary scattering will be dynamically hard.  For their encounters to have $E>0$ requires a lucky draw from the far right tail of the Maxwellian distributions that set their relative velocities, an outcome which is exponentially suppressed as one considers harder and harder binaries.

We quantify the improbability of the ``fourth outcome'' with the following idealized calculation.  First, we assume that all stars in the cluster are of the same mass $m=M_\odot$.  Next, we assume that all binaries follow Opik's law, with a semimajor axis ($a$) distribution given by 
\begin{equation}
A(a) =\left (a \ln(a_{\rm max}/a_{\rm min})\right)^{-1}.
\end{equation}
Here the minimum binary separation is set to $a_{\rm min}=3R_\odot$, and the maximum separation $a_{\rm max}$ is set to the hard-soft boundary.  If the predominant scattering events in the cluster are binary-single (binary-binary) interactions, then $a_{\rm max} = \frac{3}{2}Gm/v_{\rm rms}^2$ ($a_{\rm max} = Gm/v_{\rm rms}^2$).  The critical velocity required to render total dissociation energetically possible is 
\begin{equation}
v_{\rm 4} = \sqrt{\frac{Gm}{a_1} + \frac{Gm}{a_2}},
\end{equation}
where $a_1$ and $a_2$ are the binary semimajor axes, drawn from $A(a)$.

Next, we assume that the relative velocity of the binary-binary encounter is drawn from a Maxwellian distribution,
\begin{equation}
f_{\rm m}(v) = 6\sqrt{\frac{3}{2\pi}} \frac{v^2}{v_{\rm rms}^3} \exp\left(-\frac{3v^2}{2v_{\rm rms}^2} \right).
\end{equation}
The fraction of this distribution with velocities {\it above} a velocity $v$ is 
\begin{equation}
F_{\rm v}(v) = {\rm erfc}\left(\sqrt{\frac{3}{2}} \frac{v}{v_{\rm rms}} \right) + \sqrt{\frac{6}{\pi}} \frac{v}{v_{\rm rms}}  \exp\left(-\frac{3v^2}{2v_{\rm rms}^2} \right).
\end{equation}
Here $\rm erfc$ is the complementary error function.

\begin{figure}
\includegraphics[width=85mm]{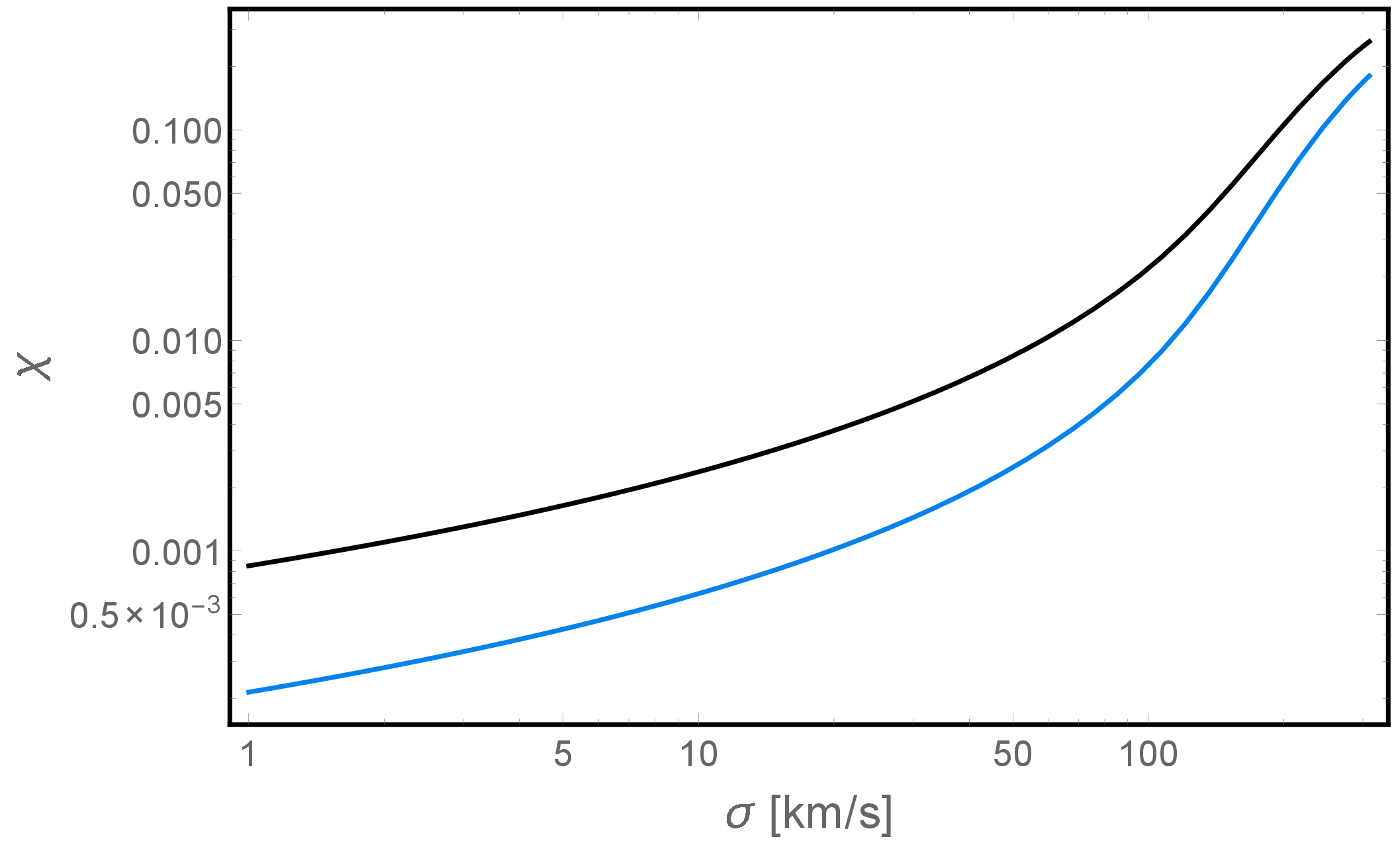}
\caption{The fraction $\chi$ of all binary-binary scatterings that make total dissociation ($1+1+1+1$) energetically possible, plotted against cluster root-mean-square velocity $v_{\rm rms}$.  In general $\chi \ll 1$, rendering the fourth outcome of binary-binary scattering quite unlikely in older collisional star systems where the hard-soft boundary sets the upper limit of the binary semimajor axis distribution.  The black line shows $\chi$ when the hard-soft boundary is set by binary-single scatterings; the blue line shows $\chi$ when the boundary is instead set by binary-binary scatterings.}
\label{fig:fig12}
\end{figure}

The fraction of binary-binary scatterings with $E > 0$ is therefore
\begin{equation}
\chi = \int^{a_{\rm max}}_{a_{\rm min}} \int^{a_{\rm max}}_{a_{\rm min}} F_{\rm m}(v_4(a_1, a_2)) A(a_1) A(a_2) {\rm d}a_1 {\rm d}a_2.
\end{equation}
This integral cannot be evaluated analytically, so we plot numerical results in Figure~\ref{fig:fig12}.  As the velocity dispersion of a star cluster increases, surviving binaries become harder to dissociate (decreasing $\chi$), but larger encounter velocities become accessible in the Maxwellian (increasing $\chi$).  The second effect dominates; $\chi$ increases with increasing $v_{\rm rms}$.  Nonetheless, $\chi \ll 1$ for realistic star clusters, justifying our focus on the three $E < 0$ outcomes of binary-binary scattering.  We note that in very young star clusters where the hard-soft boundary does not set $a_{\rm max}$ in the binary semimajor axis distribution, $1+1+1+1$ outcomes may be much more common.

\section*{Acknowledgments}

A.~M.~G. is funded by a National Science Foundation Astronomy and Astrophysics Postdoctoral Fellowship under Award No. AST-1302765.  Financial support was provided to N.~C.~S. by NASA through Einstein Postdoctoral Fellowship Award Number PF5-160145.  N.~C.~S. would also like to thank the Aspen Center for Physics for its hospitality during the completion of this research.


\bsp

\label{lastpage}

\end{document}